\documentclass{emulateapj}
\usepackage{apjfonts}
\usepackage{lscape}  
\usepackage[T1]{fontenc}
\usepackage{ae,aecompl}
\usepackage{color}
\usepackage{comment}

\newcommand{\gcc}{\ \mathrm{g\ cm^{-3} }}

\newcommand{\cms}{\ \mathrm{cm \ s^{-1}}}

\newcommand{\kms}{\ \mathrm{km \ s^{-1}}}

\newcommand{\egg}{\ \mathrm{erg \ g^{-1} }}

\newcommand{\nuclei}[2]{\ensuremath{\mathrm{^{#1}#2}}}
\newcommand{\helium}{\nuclei{4}{He}}
\newcommand{\carbon}{\nuclei{12}{C}}
\newcommand{\oxygen}{\nuclei{16}{O}}
\newcommand{\neon}{\nuclei{20}{Ne}}
\newcommand{\magnesium}{\nuclei{24}{Mg}}
\newcommand{\silicon}{\nuclei{28}{Si}}
\newcommand{\nickel}{\nuclei{56}{Ni}}

\newcommand{\iron}{\nuclei{54}{Fe}}
\newcommand{\FLASH}{{\sc flash}}

\begin{document} 

\title{Capturing the Fire: Flame Energetics and Neutronizaton for Type Ia
Supernova Simulations}

\author{
A. C. Calder\altaffilmark{1,2},
D. M. Townsley\altaffilmark{2,3},
I. R. Seitenzahl\altaffilmark{2,3},
F. Peng\altaffilmark{1,2,3},
O. E. B. Messer\altaffilmark{1,4}, \\
N. Vladimirova\altaffilmark{1},
E. F. Brown\altaffilmark{5},
J. W. Truran\altaffilmark{1,2,3,6,7},
D. Q. Lamb\altaffilmark{1,2,6}
} 

\altaffiltext{1}{Center for Astrophysical Thermonuclear Flashes,
                 The University of Chicago,
                 Chicago, IL  60637}
\altaffiltext{2}{Department of Astronomy \& Astrophysics,
                 The University of Chicago,
                 Chicago, IL  60637}
\altaffiltext{3}{Joint Institute for Nuclear Astrophysics,
                 The University of Chicago,
                 Chicago, IL  60637}
\altaffiltext{4}{National Center for Computational Sciences,
                 Oak Ridge National Laboratory,
                 Oak Ridge, TN  37831}
\altaffiltext{5}{Department of Physics and Astronomy, National Superconducting Cyclotron Laboratory, and Joint Institute for Nuclear Astrophysics,
                 Michigan State University,
                 East Lansing, MI 48824}
\altaffiltext{6}{Enrico Fermi Institute,
                 The University of Chicago,
                 Chicago, IL  60637}
\altaffiltext{7}{Physics Division,
                 Argonne National Laboratory,
                 Argonne, IL 60439}

\begin{abstract}

We develop and calibrate a realistic model flame for hydrodynamical
simulations of deflagrations in white dwarf (Type Ia) supernovae. Our
flame model builds on the advection-diffusion-reaction model of Khokhlov
and includes electron screening and Coulomb corrections to the equation
of state in a self-consistent way. We calibrate this model flame---its
energetics and timescales for energy release and neutronization---with
self-heating reaction network calculations that include both these Coulomb
effects and up-to-date weak interactions. The burned material evolves 
post-flame due to both weak interactions and hydrodynamic changes in 
density and temperature. We develop a scheme to follow the evolution, 
including neutronization, of the NSE state subsequent to the passage of
the flame front. As a result, our model flame is suitable for deflagration
simulations over a wide range of initial central densities and can
track the temperature and electron fraction of the burned material through
the explosion and into the expansion of the ejecta.

\end{abstract}
\keywords{hydrodynamics --- nuclear reactions, nucleosynthesis, abundances --- 
supernovae: general --- white dwarfs}

\section{Introduction}
\label{sec:intro}

Type Ia supernovae (SNe) are bright explosions characterized by
strong silicon P Cygni features near maximum light
and a lack of hydrogen spectral features.
The currently favored interpretation is the disruption
of a near-Chandrasekhar-mass C/O white dwarf by a thermonuclear runaway
\citep[for a review, see][and references therein]{hillebrandt+00}.
These events are fascinating in and of themselves and are important both
for their contribution to the cosmic abundance of iron-peak elements
and for their role as standard candles.

Models of Type Ia SNe necessarily involve a mechanism for
incinerating the star by a thermonuclear runaway, and the nature of this 
mechanism is the subject of contemporary research. In the explosion, a 
thermonuclear flame propagates through the C/O fuel of the white dwarf as 
either a subsonic deflagration front~\citep{nomoto+76,nomoto+84,reinecke+02,gamezo+03} 
or a supersonic detonation wave~\citep{arnett+69,boisseau+96} and releases 
sufficient energy to unbind the star.  However, models involving either a pure
deflagration or a pure detonation have traditionally been unable to 
provide an explanation for both the observed expansion velocities and the
spectra produced by ejecta that are rich in intermediate-mass and
iron-peak elements \citep{truran+71,woosley+86,woosley+90}. Recent
work, however, suggests that a fast deflagration alone may provide sufficient
energy to unbind the star~\citep{hillebrandt+05}.

There has been considerable progress recently in
hydrodynamical simulations of deflagrations of C/O white dwarfs
\citep{ropke.hillebrandt:full-star, gamezo+05} that model the entire
star. This is a complicated endeavor, predominantly due to
the vast range of length scales: the laminar flame width is $\sim
10^{-3}\textrm{--}10\,\mathrm{cm}$, some 8 to 12 orders of magnitude
smaller than the stellar radius \citep{timmes+92}. Because the
computational requirements for simulations with these disparate scales
demand resources well beyond current capabilities, multidimensional
Type Ia models must make use of an appropriate sub-grid-scale model for
the evolution of the thermonuclear burning front. Moreover, large-scale
simulations are very demanding of computational resources, and it is not
feasible at present to include enough nuclides to allow for directly
computing the reaction kinetics. A realistic model must accurately
describe the nuclear energy that is released, the timescale on which
it is released, and the compositional changes that occur in the flame.
In addition, the burned material continues to evolve after the passage
of the flame due to both weak interactions and hydrodynamic evolution,
and realistic simulations must describe this ``post-flame" evolution.

In this paper we present a study of the nuclear burning that occurs
during C/O deflagrations, with the goal of  producing a realistic
flame model for simulations of Type Ia supernovae. Building on the
advection-diffusion-reaction (ADR) flame model of~\citet{khokhlov95}, we
further develop a three-stage flame model. The
work described here improves upon earlier calculations in
several ways. First, screening and the Coulomb interaction are included
self-consistently in the forward and inverse rates, so that detailed
balance is preserved. Second, we account for the neutronization and
neutrino loss rates via electron captures at high densities. Our treatment
is sufficiently general that it can also describe the expansion following
the explosion. With minimal tuning, the method should capture the bulk
energetics and associated ``freezing" of the abundance pattern as the
matter expands.

The present paper is concerned with describing with the utmost care the
nuclear processes that occur in the real flame and proposing a method that
can capture the necessary features: energy release, its timescales and
neutronization.  The hydrodynamic behavior in the 3-dimensional flow (importance
of flame front curvature, acoustic behavior, flame front stability, effects of
finite resolution, etc.) is under active investigation.  Due to the
complexity of those issues, the fact that the work is still in progress, and
the clean separation of much of them from the nuclear physics addressed here,
we have chosen to publish this work in advance of a forthcoming separate
hydrodynamical study. It is important to note that such a hydrodynamical
study must be performed in order to understand how this particular
flame-capturing technique, with realistic energetics, behaves in simulations
of a deflagration in a WD.

Our three-stage flame model describes the propagation of a nuclear flame
through a uniform \carbon/\oxygen\ mixture.  In this work we consider
the case of a mixture with 1:1 mass fractions, but
the method can be applied to any ratio. As described
in \citet{khokhlov+01}, the burning occurs in roughly three stages with
well-defined timescales. First is the $\carbon+\carbon$ fusion, leading
to a mixture consisting of the unburned \oxygen\ together with \neon,
\magnesium, and $\alpha$-particles.  Second, the resulting mix burns on
a longer timescale leading to the formation of predominantly Si-group (intermediate
mass) $\alpha$-elements and $\alpha$-particles, a state
that is commonly referred to as nuclear statistical quasi-equilibrium
(NSQE). Third, these elements burn on a still longer timescale to nuclear
statistical equilibrium (NSE), which at the densities of interest in the
deflagration phase of Type Ia SNe, consists primarily of Fe-peak nuclei,
$\alpha$-particles, and protons.

After the flame has burned the C/O fuel, two factors continue to influence 
the energetics: (1) decreases (or possibly increases) in temperature 
and density resulting from the large-scale motion of the star, and  
(2) neutronization of the hot nuclear ash. Neutronization can heat or cool 
the ash, depending on whether the nuclear binding energy released by 
neutronization exceeds or falls short of the energy lost by neutrinos, and  
the competition between these two processes is a sensitive function of density.  
We note that because the light curves of SNe Ia are strongly sensitive to 
the mass of $^{56}$Ni produced, the degree to which neutronization reduces 
the mean $^{56}$Ni concentration over the regions of the core (extending out 
to $\approx$ 0.8-1 M$_\odot$) in which conditions of nuclear statistical 
equilibrium are achieved is clearly a critical issue.  It is therefore important 
to provide an accurate measure of the influence of electron captures on the 
energetics. Only in this way can we ensure that the peak temperatures and 
temperature histories of fluid elements are accurate. Accordingly, our flame 
model follows this neutronization. 

In this paper we describe
nuclear flame propagation (\S~\ref{sec:flame_propagation}). We describe fully-resolved simulations of thermonuclear 
flames propagating through a slab of stellar material and compare these to similar
``one-zone" simulations of nuclear burning. The ``one-zone" or ``self-heating" simulations 
allow for a longer-time evolution than the fully-resolved flames and provide the energetics
and timescales for the model flame. Comparison to the fully-resolved flame calculations 
is an important consistency test. In this section we also describe the method
by which we determine the density contrast across the flame as is required as
input by the flame model and present contrasts at selected densities measured
from the fully-resolved flame simulations. In (\S~\ref{sec:flame}) we describe
our model flame. We outline the basics of our ADR
flame capturing scheme and describe the stages of nuclear burning and the energy
release occurring in each. In (\S~\ref{sec:network}), we describe the calculations
of energetics and timescales that serve as input to the flame model. We present
the results of self-heating simulations of nuclear burning with a detailed nuclear 
reaction network and
contemporary reaction rates. We report on the effect of including Coulomb corrections
and screening and present timescales for burning to NSQE and NSE. In 
(\S~\ref{sec:postflame}), we describe the change in energy resulting from the
evolution of the burned material in NSE. We present a study of the NSE state
that quantifies the change in binding energy with temperature and density evolution
and the effects of weak interaction (neutronization). We also describe our method
of representing the NSE state with a reduced set of nuclei. In (\S~\ref{sec:results})
we present results of model flame simulations at several densities, and in 
(\S~\ref{sec:conclusions}) we draw conclusions from this effort. Finally, we present
an appendix with details of the plasma coulomb corrections applied to both the
network and NSE calculations.

\section{Nuclear Flame Propagation}
\label{sec:flame_propagation}

A thermonuclear flame in the interior of a star propagates subsonically. 
This slow propagation makes following the flame with an explicit hydrodynamics 
method such as PPM difficult because of the high spatial resolution needed
to resolve the flame. The time step in an explicit hydrodynamics 
method is limited by the Courant condition; namely, the code must resolve 
the sound-crossing time of the smallest zone. Because of this constraint,
the time steps for a high-resolution simulation are very small and a large 
number of time steps is therefore required to propagate the subsonic flame,
even across just a few simulation zones. 
This problem is particularly acute at lower densities for which the timescale 
for burning is relatively long. A fully-resolved flame simulation could literally
take billions of time steps, which is impractical and would most likely produce
inaccurate results. We implement the flame model in the explicit hydrodynamics 
code \FLASH\ \citep{fryxell+00,calder+02}, but rely for input to the model flame on 
self-heating network calculations described in this section.

\subsection{Nuclear Flame Structure and Self-Heating Calculations}

Because of the difficulty encountered with propagating flames with \FLASH,
particularly at low densities, we made use of self-heating network
calculations (see section \ref{sec:network} for details).  These are
(energetically) closed-box calculations of nuclear burning at either constant
pressure (isobaric) or constant density (isochoric).  We use both cases in
this paper, each for a different purpose.  Isobaric behavior should be quite
close to what occurs in a fluid element as the subsonic flame sweeps through
it, so that, in a sense, such a self-heating network calculation may be
thought of as a ``one-zone" calculation.  Isochoric calculations are
thermodynamically simpler and for this reason are used in section
\ref{sec:network} to gauge the effects of screening and to compare different reaction networks.
As will be shown below, the results of
isobaric and isochoric calculations are fairly similar for high density
($\rho > 10^8$ g cm$^{-3}$).

\begin{figure}
  \includegraphics[width=\hsize]{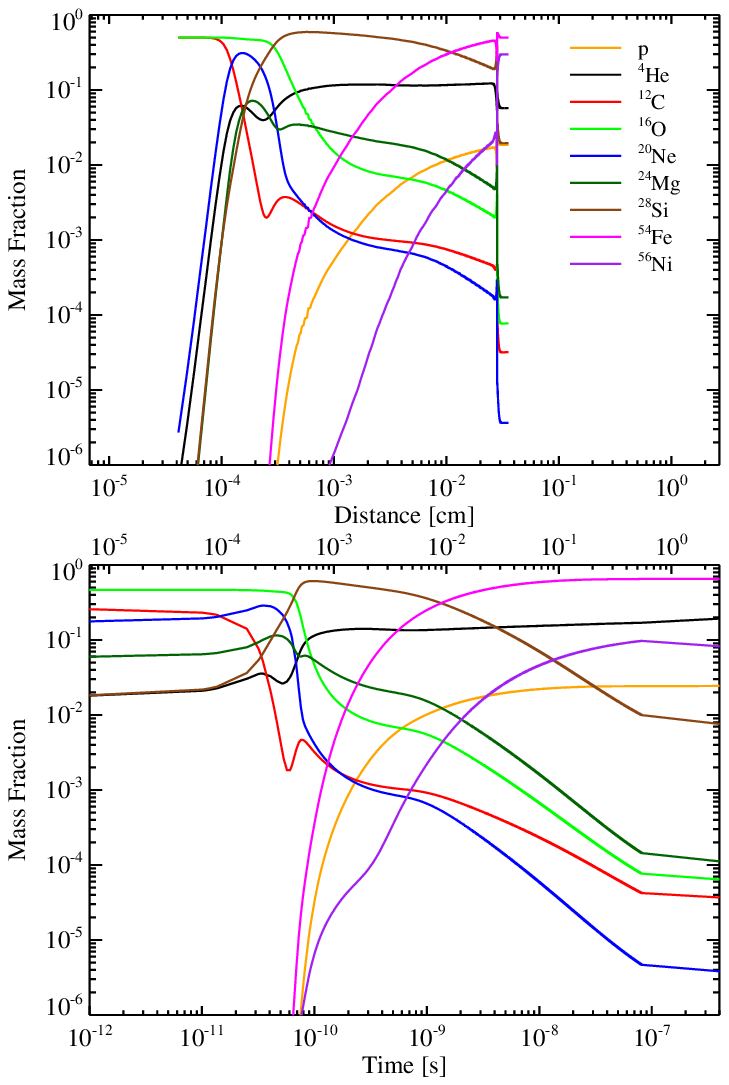}
\caption{Top panel: Results from a fully resolved DNS of a propagating flame 
   at $10^9 \gcc$ performed with \FLASH\ and the \textit{aprox19} network. 
   Shown are abundances for p, \helium, \carbon, \oxygen, \neon, \magnesium, 
   \silicon, \iron, and \nickel\ 
   plotted against the distance from the flame position so that the flame is propagating
   to the left. The small ``tails" 
   at the right are artifacts of the ``match head" 
   used to ignite the flame. The simulation was run on an adaptive mesh
with an effective resolution of $7.81 \times 10^{-6} \mathrm{cm}$.
\label{fig:isobar}
Bottom panel:
Results from an isochoric simulation 
   at $10^9 \gcc$ performed with the self-heating 19 nuclide 
network \textit{aprox19}. 
Shown are the same abundances as the top panel
plotted against the time and the 
calculated distance. The distance scaling along the top axis was obtained 
by multiplying the time by the flame speed with respect to the (expanded) ash.}
\label{fig:isochor}
\end{figure}

\begin{figure}
  \includegraphics[width=\hsize]{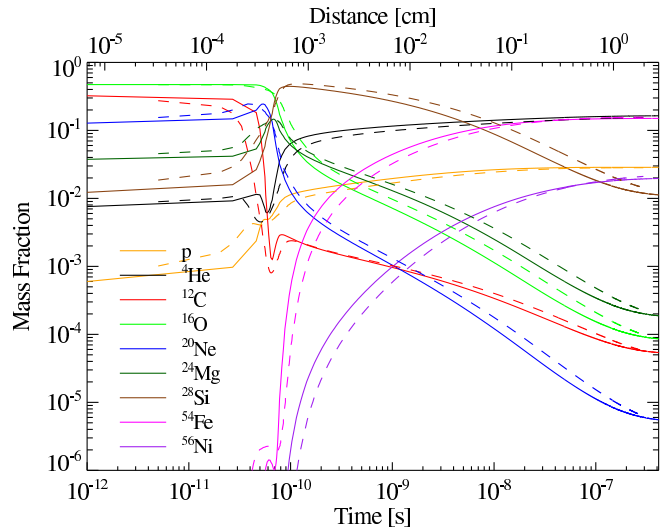}
 \caption{Abundance evolution of burning an equal-mass mix of C and O from an
initial density of $10^{9}\gcc$. Shown are the results from a isochoric
(\emph{solid lines}) and an isobaric
($P=5\times 10^{26}$ dyne~cm$^{-2}$; \emph{dashed lines}) self-heating
calculation.  Our 200 nuclide network was used in both calculations.  The
approximate distance scale is as in Fig.~\ref{fig:isochor}.}
\label{fig:abund_r=1e9}
\end{figure}

We refer to an actual propagating flame simulated with \FLASH\ using thermal
diffusion as a direct numerical simulation (DNS).  An important test is the
consistency between actual flames propagated with \FLASH\ and the results of
the self-heating calculations. We performed simulations of propagating flames
with \FLASH\ at several relatively high densities for which a flame
calculation was feasible.  The most notable difference between the DNS and
the self-heating calculations is that it is necessary to use a fairly small
reaction network (19 nuclides) in the DNS, whereas the self-heating
calculations are able to use a much larger network (200 nuclides, see section
\ref{sec:network}). The DNS employed adaptive mesh refinement with an effective 
resolution of $7.81 \times 10^{-6} \mathrm{cm}$. The minimum refinement of the mesh 
was kept at $1.56\times 10^{-5} \mathrm{cm}$, and one additional level of AMR
allowed the extra resolution of the flame front.

A comparison of a DNS and isochoric self-heating is
shown in Fig.~\ref{fig:isobar}.  Both of these calculations use the
19 nuclide network \textit{aprox19}~\citep{wzw1978,timmes+99} to burn fuel
which is initially an equal mass mixture of \carbon\ and \oxygen\ at a
density of $10^9\gcc$. The top panel shows how abundances vary with distance
behind the flame front in the DNS (so that unburned material is on the left).
The flame was ignited at the edge of a 0.08 cm domain by raising the
temperature of a small region of the domain to the unphysically high value of
$10^{10}$K. Heat then diffuses from this preheated region and ignites the
flame, which then propagates across the grid.  The rightmost parts of the
mass fraction curves, at the largest distance from the flame, have small
``tails'' that result from the unphysical initial conditions of the very hot
``match head.'' Shown are mass fractions for p, \helium, \carbon, \oxygen,
\neon, \magnesium, \silicon, \iron, and \nickel.  The flame speed calculated
from this simulation, $34.2 \kms$, agreed well with the result
of~\citet{timmes+92}.

The abundance evolution with time from an isochoric self-heating calculation
using the same network is shown in the bottom panel of
Fig.~\ref{fig:isobar}.
We have calculated an approximate distance from the flame front (shown on the
upper scale) by
multiplying the time by the flame speed with respect to the ash.
From conservation of mass this is related to the laminar flame speed in the fuel by 
$S_{\rm ash} =  S_{\rm lam} (1+A)/(1-A)$, where 
$A = (\rho_{\rm fuel} - \rho_{\rm ash})/(\rho_{\rm fuel} + \rho_{\rm ash})$ 
is the Atwood number.  For this calculation we used values of $A_{\rm P200}$ in 
Table~\ref{tab:atwood} (see description in \S~\ref{sec:atwood})
and the laminar flame speed from \citet{timmes+92}, 
$S_{\rm lam} = 5.3 \times 10^6~{\rm cm\,s^{-1}}$ for 
$\rho = 10^9\gcc$ and $X(\carbon) = 0.5$ (see eq.~[\ref{eq:laminar}]). 
We find good agreement in terms of abundance structure between the isochoric self-heating
calculation and DNS.  The stretching on the left in the bottom
panel is due to the choice of zero point. With an initial temperature
$T=1.7\times10^{9}~{\rm K}$, there is a long smoldering phase in the
self-heating calculation before rapid consumption of \carbon\ occurs.
The same "pre"-heating is accomplished in the DNS by thermal diffusion,
giving a heating time and length scale similar to the \carbon\ consumption
scales.
To be able to broaden the region where C/O decreases sharply, we
chose the time zero to be a time where C/O starts
to decrease.  Since the C consumption timescale is short compared to the
later flame stages (e.g. reaching NSE), this choice does not affect the total
flame thickness.
Also the depletions of \carbon\ and \oxygen\ are
shifted slightly further in distance in the self-heating plot by our use of a constant
distance scaling because this is in reality where the expansion takes place.

The next step is to compare these results with those of the much larger nuclear network
that will be used for the rest of our self-heating calculations.  Fig.~\ref{fig:abund_r=1e9} shows the abundance
evolution for $\rho=10^9$ g cm$^{-3}$ using the 200 nuclide network described
in \S~\ref{sec:network}.  The
two networks give very similar evolutions,  with the most notable difference
being in the final abundances.  The 200 nuclide network has many more
nuclides available near the Fe peak, such that less material is concentrated
in a single nuclide ($^{54}$Fe) than with the small network.  The 5 most
abundant nuclides in the NSE state are
$^{54}$Fe, $^4$He, $^{55}$Co, $^{58}$Ni, $^{56}$Co with abundances of 16\%,
13\%, 8.4\%, 7.3\% and 6.2\% respectively, and there are another 14 nuclides
with abundances between 1 and 5\%.

The evolution of a fluid element as the subsonic flame passes is in fact
nearly isobaric.  Due to the high level of degeneracy at $\rho=10^9$
g cm$^{-3}$, however, there is little expansion and so the isochoric and isobaric
self-heating calculations give quite similar results.  In
Fig.~\ref{fig:abund_r=1e9}, we also show the result from an isobaric
calculation. Here the pressure ($P=5\times 10^{26}~{\rm dynes\,cm^{-2}}$) is
chosen to give an initial density of $10^9$ g cm$^{-3}$. We can see that the
isobaric calculation takes slightly longer to reach NSQE, a
result of a slightly decreased temperature due to expansion. The density is
decreased to $8.3 \times 10^9\gcc$ at this phase. However, the NSE timescale
and the final NSE abundances don't differ much from the isochoric
calculation.


\begin{figure}
  \includegraphics[width=\hsize]{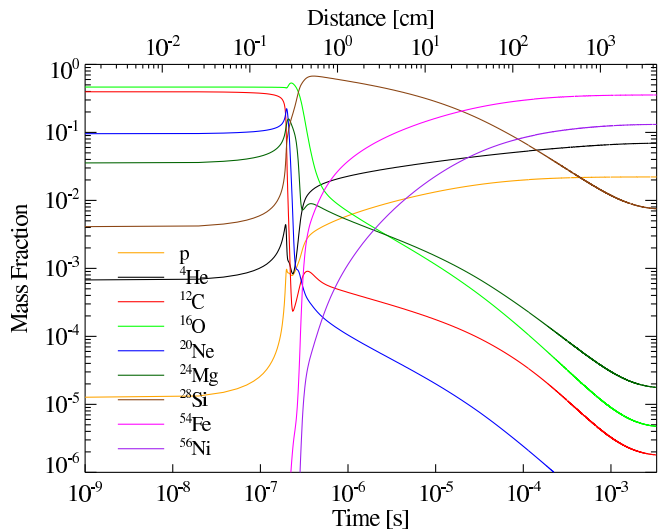}
\caption{Similar to Fig.~\ref{fig:abund_r=1e9} but for isobaric only with
an initial density of $10^{8}\gcc$. }
\label{fig:abund_r=1e8}
\end{figure}
\begin{figure}
  \includegraphics[width=\hsize]{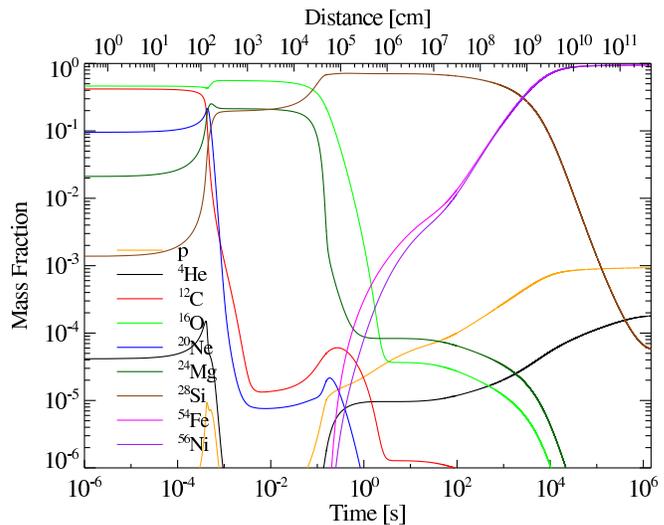}
\caption{Similar to Fig.~\ref{fig:abund_r=1e9} but for isobaric only with
an initial density of $10^{7}\gcc$.}
\label{fig:abund_r=1e7}
\end{figure}
Before discussing the general features of this flame structure, it is useful
to see how it changes when burning at lower densities.  Figs.
\ref{fig:abund_r=1e8} and \ref{fig:abund_r=1e7} show isobaric self-heating
calculations for initial densities of $\rho_i = 10^8$ and $10^7$ g cm$^{-3}$.
Light particles, particularly $\alpha$-particles, become less dominant in the
NSE final state.
It is already evident that at these low densities the flame becomes very
extended physically, such that a profile like that shown in
Fig.~\ref{fig:abund_r=1e7} could never be realized in an actual star.
The dynamical response time of the star is shorter than these
timescales (see Fig.~\ref{fig:screen_time}), causing the reactions to freeze out due 
to expansion, and the scale height at this density is smaller than the large scales 
of the structure shown.

At all densities, though the overall progression timescale differs
significantly, the three burning stages are evident.  Beginning with equal
parts by mass of $^{12}$C and $^{16}$O, initially the burning of the $^{12}$C
leads to a mixture of $^{16}$O, $^{20}$Ne, $^{24}$Mg, and $\alpha$-particles.
This mixture then burns to intermediate mass ($^{28}$Si to $^{40}$Ca) nuclei
and $\alpha$-particles
(NSQE) in which most of the energy release has taken place.  Finally, the
mixture reaches full NSE consisting of a temperature and density dependent
mixture of iron-peak nuclei, $\alpha$-particles, and protons.  The transition
points at $\rho=10^8$ g cm$^{-3}$ (Fig. \ref{fig:abund_r=1e8}) are at
0.2, 0.4, and 30 cm on the approximate
distance scale.  While the $^{12}$C burning and the transition to NSQE are
well separated at the low density, at the higher densities these stages
appear more as an offset between consumption of $^{12}$C and $^{16}$O.

It is important to note again that the NSE state is not static.
The forward and inverse rates are large but balanced, which enables the
NSE state to adjust as the fluid element evolves under hydrodynamical
motions (see \S~\ref{sec:postflame} for further details).  The flame
model must reproduce this behavior with good approximations to the flame
propagation speed, energy release timescales, and total energy release.

\subsection{Inferring Density Contrast From Isochoric Calculations}
\label{sec:atwood}

In the flamelet regime, the flame is accelerated by the Rayleigh-Taylor
instability, which increases the surface area of the flame and its effective
propagation speed \citep{khokhlov95,niemeyer+kerstein97}.  The strength of
this effect is related to the density contrast, characterized via the Atwood
number, $A$, due to the energy release in nuclear burning.  Energy is
released and does work to expand the matter and raises the temperature until
NSE is reached (for densities $\gtrsim 10^7\gcc$) or the nuclear energy is
exhausted upon fusion to \nickel.  

The energy release found from an isochoric self-heating calculation can be
used to estimate the expansion expected for the actual (approximately
isobaric) burn.  We denote this approximate Atwood number by $\tilde A$, and
calculate it by using a single-stage ADR flame (see Section \ref{sec:adr})
that has a specified energy release, $\Delta Q$, and ash composition taken from
an isochoric self-heating calculation (these are given in Table
\ref{tab:big}).  The ash composition defines both the heat capacity and the
relationship between the ion pressure contribution and the temperature.
Table~\ref{tab:atwood} compares Atwood numbers calculated using
high-resolution DNS of flames, $A_{\rm DNS}$, as was shown in
Fig.~\ref{fig:isobar}, and $\tilde A$ for isochoric self-heating
calculations computed with two reaction networks, {\it aprox19} and P200.
By comparing these three cases we can see what differences are due to the
nuclear network alone, and what are due to the isochoric
approximation.  Self-heating calculations with the {\it aprox19} nuclear
network, the same as that used in the DNS, give a larger total energy release
at the high densities relative to those found using our 200 nuclide network (see
column 5), which is reflected in $\tilde A$ (column 3 and 4).

The approximate Atwood numbers, $\tilde A$, are lower than the actual Atwood
numbers.  The ash temperatures, $\tilde T_{\rm ash}$, found for the P200 case
in this approximate scheme are shown in column 6, and are lower than the
final temperatures (shown in Table~\ref{tab:big}) of the self heating
calculations from which the $\Delta Q$ was drawn, reflecting the fact that
some of the energy was used to do work for the expansion.  This lower
temperature means that the ash is now out of NSE, and if allowed to relax it
would release nuclear binding energy, expand, and raise its temperature to
reach the $A$ found in the DNS.

We find that, at these densities ($\rho>10^8$ g cm$^{-3}$), isochoric
self-heating calculations provide reasonable estimates for the
expansion factor and very gross estimates for the energy release.
From this we conclude that $\tilde A$ from the isochoric self-heating
calculation can be used as a static approximation where needed.  We should
emphasize, however, that in the model flame described below, the actual total
energy release of the flame is determined by the NSE in post-flame material,
which is calculated dynamically as described in section \ref{sec:postflame},
and therefore will reproduce that of an isobaric flame.

\begin{deluxetable*}{ccccccc}
\tabletypesize{\footnotesize}
\tablewidth{0pc}
\tablecaption{Atwood numbers\label{tab:atwood}}
\tablehead{ \colhead{density} & \colhead{$A_{\rm DNS}$} & \colhead{$T_{\rm ash \; DNS}$} & \colhead{$\tilde
A_{\rm aprox19}$} & \colhead{$\tilde A_{\rm P200}$} & \colhead{$\Delta Q_{\rm aprox19} - \Delta Q_{\rm P200}$}
&  \colhead{$\tilde T_{\rm ash}$}  \\
\colhead{($10^9\gcc$)} &\colhead{}  &    \colhead{$10^9\gcc$)}  & \colhead{ } & \colhead {}  &  \colhead{(10$^{17}\egg$)}   &  \colhead{(10$^9$ K)}}
\startdata
6    & 0.065 & 10.5 &   0.056 &  0.051 & 0.21 &  6.7 \\
2    & 0.093 & 9.34 &   0.082 &  0.077 & 0.15 &  6.8 \\
1    & 0.116 & 8.51 &   0.101 &  0.097 & 0.08 &  6.6 \\
0.5  & 0.150 & 7.80 &  0.127  & 0.125  & 0.05 &  6.3 \\
0.1  &      & 	    &  0.227 &  0.229  & -0.06&  5.3 \\
\enddata
\end{deluxetable*}

\section{Implementation of Model Flame Energetics}
\label{sec:flame}

The key ingredient in numerical 
simulations of the deflagration phase of SNe Ia is the nuclear flame model.
In this section we describe our flame model, the implementation of
the energetics, and the implementation of the ``post flame" NSE
evolution.

\subsection{Basics of an ADR Flame model}
\label{sec:adr}

\begin{figure}[b]
\includegraphics[width=\hsize]{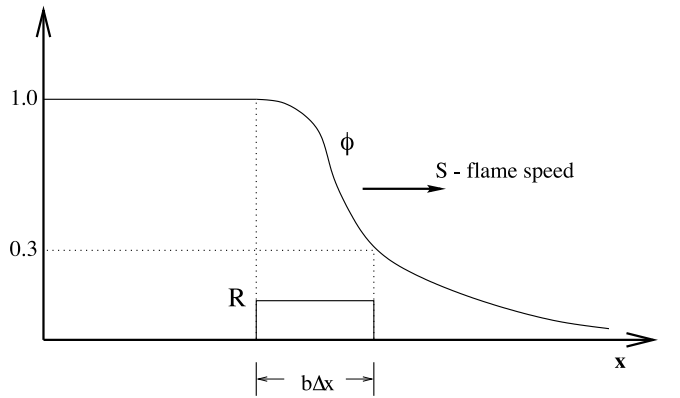}
\caption{{flame front structure for a one-stage ADR flame}}
\label{fig:flame1}
\end{figure}
We use a flame-capturing scheme based on the ADR method
\citep{khokhlov95,khokhlov+01} that is described in
detail in~\citet{vladimirova+05}.  The flame
front is localized via the value of a reaction progress variable,
$\phi$, such that in the reactant $\phi = 0$, in the product $\phi = 1$, and $\phi$ varies monotonically across the
flame front (see Fig.~\ref{fig:flame1}). The width of the ADR flame is often much
larger than the thickness of the physical flame it represents, as it
is prescribed to be several computational zones thick, thereby
becoming even kilometers in width in full-star supernova simulations.
The value of $\phi$ is defined everywhere on the grid, and can be
associated with the mass fraction of burned material in a cell, provided that the reactant composition is initially uniform.

The reaction progress variable is evolved via an 
advection-diffusion-reaction equation,
\begin{equation} \label{eq:adr}
{{\partial \phi}\over{\partial t}} + {\bf v}\cdot\nabla \phi
= \kappa \nabla^2\phi + \frac{1}{\tau}R(\phi)~, \end{equation}
with artificial reaction and diffusion coefficients
\begin{equation} \label{eqS7}
\kappa	=  {\rm const},\qquad
R(\phi) = \left\{ \begin{array}{ll}
      R_0 ={\rm const.,} & {\rm if}~~ \phi_0 \leq \phi \leq 1 \\
      0,	      & {\rm otherwise,}
   \end{array}	\right.	 \\
\end{equation}
where $\phi_0$ marks the value of $\phi$ at which the reaction begins.

When $\bf{v}=0$, the solution of equation (\ref{eq:adr}) is a
traveling wave with speed $S_0$ with a specific functional form, 
$\phi(x-S_0t)$.  Traveling speed $S_0$ and front thickness $l_0$ 
depend on the diffusivity, the reaction time, and the amplitude of the
reaction rate, $S_0 \propto \sqrt{R_0\kappa/\tau}$ and $l_0 \propto
\sqrt{\kappa\tau/R_0}$.  For the reaction rate (\ref{eqS7}), the expressions
for $S_0$ and $\phi(x-S_0t)$ can be found analytically.  As shown in
\cite{vladimirova+05}, if the amplitude of the reaction rate $R_0$
satisfies the relation $(\phi_0 - R_0)e^{1/R_0} + R_0 = 0$, the front 
propagates with the speed $S_0=\sqrt{\kappa/\tau}$.  In our
implementation, $\phi_0= 0.3$ and $R_0 = 0.3128$, so the front speed
and thickness obey the relations $S_0=\sqrt{\kappa/\tau}$ and
$l_0 = 4\sqrt{\kappa\tau}$.

When $\bf{v}
\ne 0$, the front propagates with the speed $S_0$ with respect to the
background velocity, provided the velocity variations on the scale of
the front thickness are negligible in comparison with $S_0$.  

The ADR flame model uses the evolution of $\phi$ according to
equation~(\ref{eq:adr}) to propagate a front.  The front speed $S$ and
the front thickness $l$ are input parameters; they determine the
coefficients $\kappa$ and $\tau$ in equation~(\ref{eq:adr}),
\begin{equation}
  \label{eq:ck_new} \kappa = aSl/4, \qquad
     \tau = l /(4aS).
\end{equation}
Here, $a$ is an adjustment coefficient that accounts for the front
speed dependence on thermal expansion (since thermal expansion results
in velocity variations across the front and alters the front speed). The
coefficient $a$ depends on the density ratio $\alpha = \rho_{\mathrm u} /
\rho_{\mathrm b}$ and, for the reaction rate (\ref{eqS7}), was estimated as
$a = 1+0.3(\alpha-1)\approx 1+0.6A$~\citep{vladimirova+05}.  Here
$\rho_{\mathrm u}$ and $\rho_{\mathrm b}$ are the unburned and burned
densities, respectively. 
Note that when $a=1$, we have $S=S_0$ and $l=l_0$.  An empirical calibration
was made based upon flames propagated with the multistage flame (described
below), finding $a\approx 1+1.5A$.  We attribute the difference to the energy
release occurring in the second stage, slightly after the ADR flame front.  The
front thickness is set to be several grid points (spaced by $\Delta x$) per
interface $l=b\Delta x$, usually with $b=4$.  Further details of the
implementation of this flame-capturing method may be found
in~\citet{vladimirova+05}.

The above description of the ADR flame capturing scheme describes the
method of propagating a model flame at a specified speed. Flames passing through stellar material are subject to fluid instabilities, particularly the Rayleigh-Taylor instability, which increases mixing and
therefore the effective flame speed.  Analyzing the balance between generation and destruction of flame
surface in a steady turbulent burning regime, \citet{khokhlov95} showed that the flame surface packed in a given volume is inversely proportional to the laminar flame speed.  Consequently, the total burning rate, which is equal to the product of the laminar flame speed and the flame surface area, is independent of the laminar flame speed in this regime. Numerical calculations of flames propagating in a vertical column demonstrate that the effective flame speed is determined by the
large-scale motion rather than by the local flame properties
\citep{khokhlov95,zhang+06}.  To account for flame acceleration on scales unresolved by the
simulation, we use a turbulent subgrid flame speed \citep{khokhlov95},
\begin{equation}
 \label{eq:turb}
     S_{\rm sub} = 0.5 \sqrt{A g \Delta x},
\end{equation}
where $g$ is the local gravitational acceleration. As an input to ADR model
we take $S = \max (S_{\rm lam}, S_{\rm sub})$, where the laminar flame
speed,
\begin{equation}
 \label{eq:laminar}
     S_{\rm lam} = 9.2\times10^6 \,
		  \left(\frac{\rho_{\mathrm u}}{2\times 10^9\gcc  }\right)^{0.803}
		  \cms,
\end{equation}
is from \citet{timmes+92}.

The subgrid flame speed~(eq.~[\ref{eq:turb}]) depends on the Atwood number $A$. 
To compute $A$ exactly, one must know the densities ahead of and behind the
flame interface. The straightforward approach of ``taking a probe'' on
both sides of the flame works well with models in which an
infinitely-thin interface is reconstructed inside the computational
cell and information about burned and unburned fluid is available in
every partially burned cell. In the ADR flame model, to compute $A$ in the partially burned cell, one must ``take a probe''
at the location several computational cells away.  It is not easy to
develop a robust algorithm for locating pure fuel and pure ash in the
vicinity of a particular partially burned cell. Efficiently implementing
such an algorithm in a parallel, domain-decomposed simulation code such as
\FLASH\ is even more complicated because the information needed to
perform the calculation for a given computational zone may not be
directly available.

An alternate approach is to use {\em local} information to estimate
the burned and unburned states from the partially burned state \citep{khokhlov95,vladimirova+05}.  The approach is based on
the proportionality between the mass fraction of product $\phi$ and the
fraction of energy released, so that the thermodynamical state can be
parameterized by $\phi$. In particular, the density of the partially
burned fluid can be obtained from the unburned state and $\phi$,
\begin{equation} \label{rhophi}
    \rho = \frac{\rho_{\mathrm u}}{1+(\alpha -1)\phi},
  \qquad
    \alpha = 1 + \frac{\gamma -1}{\gamma} \frac{q \rho_{\mathrm u}}{p_{\mathrm u}}.
\end{equation}
Here, as above, $\alpha =\rho_{\mathrm u}/\rho_{\mathrm b}$ is the density ratio across the 
interface and $q$ is the mass-specific energy released in the flame. We approximate the EOS with $\gamma$, defined by $P = (\gamma - 1)\rho\mathcal E$, where $\mathcal E$ is the specific internal energy.  
The above relations assume a very subsonic flame, but $A$ is not required to be small.  Under these conditions, the pressure variation across the flame is insignificant, $p \approx p_{\mathrm u}$, and equations~(\ref{rhophi})
can be solved for $\alpha$ and $\rho_{\mathrm u}$,
\begin{equation}\label{rho_u}
     \rho_{\mathrm u} = \frac{\rho}{1-\epsilon \phi}, \quad
     \alpha = 1 + \frac{\epsilon}{1-\epsilon \phi}, \quad
   \mathrm{where} \quad
     \epsilon \equiv \frac{\gamma - 1}{\gamma}\frac{q\rho}{p}.
\end{equation}
Then the Atwood number $A=(\alpha-1)/(\alpha+1)$ and the unburned
density $\rho_{\mathrm{u}}$ needed to evaluate the flame speed may be computed using
eq.~(\ref{rho_u}).  For small Atwood numbers, we note $A\approx \epsilon/2$,
which serves as a phi-independent approximation. Similarly, $S_{\mathrm{lam}}$ is 
calculated using the local density rather than $\rho_{\mathrm{u}}$.  This implementation 
has been tested for a variety of conditions and was found to perform quite well in 
situations without extreme shear.


\subsection{The Stages of Nuclear Burning and Energy Release in the
Model Flame}
\label{sec:stages}

\begin{figure}
\includegraphics[width=\hsize]{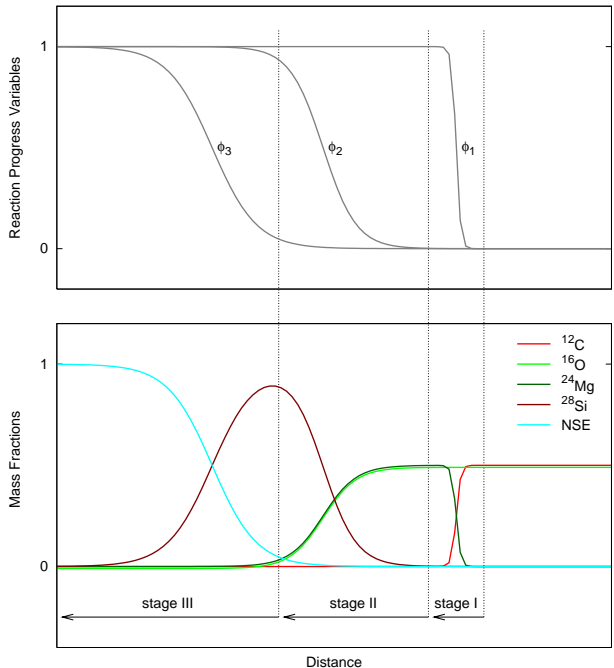}
\caption{{Diagram of three-stage model flame front structure for a flame
propagating to the right.  The upper panel shows the progress variables used
to track the flame activity:  $\phi_1$ models the \carbon\ flame front,
$\phi_2$ models progress to NSQE, and $\phi_3$ models relaxation to NSE.
The lower panel demonstrates the elements used to represent the stages and
their relation to the progress variables.}
\label{fig:3stage}}
\end{figure}
As mentioned in section \ref{sec:flame_propagation}, the structure of the
nuclear flame can be modeled as a three-stage process in which each
consecutive stage occurs on a longer timescale.  In our model, the
progress of each of these stages is tracked with progress variables $\phi_1$,
$\phi_2$, and $\phi_3$, which evolve from 0 to 1.  This is shown
schematically in the upper panel of Fig. \ref{fig:3stage}, where the flame
is propagating to the right.  Material begins as a mixture of \carbon\ and
\oxygen.  In stage 1, the \carbon\ is consumed, in stage 2, the resulting ash
burns to NSQE, and finally in stage 3
material relaxes to actual NSE.  The NSQE state is reached much more quickly
than the final NSE state, and most ($> 90\%$) of the energy release in
burning to NSE is released upon reaching NSQE \citep{khokhlov95}.  Having
separate progress variables for NSQE and NSE allows the energy release
timescale to be properly modeled while still tracking when material will
reach actual NSE abundances.  The resulting NSE state is not static, but can
change as the material evolves hydrodynamically (e.g.\ decompression in a
rising bubble), and due to weak interactions.  The treatment of this
post-flame NSE evolution is described in section \ref{sec:postflame}.

Because the burning of \carbon\ occurs on much less than the grid scale,
and the physical scale of this flame front is an important factor in setting
its propagation speed, the ADR flame scheme is used to evolve $\phi_1$ with
the parameterized flame speed discussed above.  This reflects the propagation
of the actual flame front through the star.  Although the consecutive stages
separate cleanly on a microscopic level, as indicated in figures
\ref{fig:abund_r=1e9}-\ref{fig:abund_r=1e7}, in order to allow for proper
advection of the unresolved transition fronts in the model, all three stages
are allowed to be partially progressed in the same cell.  However, later
flame stages only progress when previous flame stages have completed.  When
$\phi_1 <1$, $\dot\phi_2=0$, but behind the propagating \carbon\ flame, when
$\phi_1=1$,
\begin{equation}
\dot\phi_2 = \frac{1-\phi_2}{\tau_{\rm NSQE}}.
\end{equation}
Here $\tau_{\rm NSQE}$ is the timescale for reaching the NSQE state
evaluated in nuclear network studies discussed in detail in the following
section.  The speed of this evolution and that for $\phi_3$ below are limited
such that $\phi_2$ will change by at most 2\% in a single explicit
hydrodynamic timestep, in order to reduce noise.

As noted earlier, the bulk of the energy released in the burn
to NSE is released upon reaching NSQE. Accordingly, and for 
simplicity, all the energy available in the initial burn to NSE is released in
the first two stages, so that the nuclear energy release in the flame
is given by
\begin{equation}
\epsilon_{\rm nuc} = 
\left\{
\begin{array}{ll}
\dot\phi_1\Delta Q_{\rm C}
+\dot\phi_2(\Delta Q_{NSE}-\Delta Q_{\rm C})
& \text{when $\Delta Q_{\rm NSE } > \Delta Q_{\rm C}$,}\\
\\
\dot\phi_1\Delta Q_{\rm NSE}
& \text{when $\Delta Q_{\rm NSE} < \Delta Q_{\rm C}$.}\\
\end{array}
\right.\label{eq:flameenergy}
\end{equation}
Here $\Delta Q_{\rm C}$ is the energy release for burning the initial
abundance of \carbon\ to \magnesium\, and $\Delta Q_{\rm NSE}$ is the energy
released for burning to NSE.  $\Delta Q_{\rm NSE}$ is a function of the local
density, and is evaluated in detail in the next section.
Even at high densities, $\Delta Q_{\rm NSE} > 3\times
10^{17}\,\mathrm{ergs\, g^{-1}}$ (see Fig.~\ref{fig:screen_energy}).  The \carbon\
burning releases $\Delta Q_{\rm C} = 2.78\times 10^{17}\,\mathrm{ergs\, g^{-1}}$ for
$X_{\rm C}=0.5$, and, accordingly, the second form in equation 
(\ref{eq:flameenergy}) is not used in the simulations presented in this work.

Finally, after NSQE is reached, the final progress variable is evolved.  When
$\phi_2=1$,
\begin{equation}
\dot\phi_3 = \frac{1-\phi_3}{\tau_{\rm NSE}},
\end{equation}
where $\tau_{\rm NSE}$, evaluated below, is the timescale on which full
relaxation to the NSE abundance distribution occurs.  The evolution of the
NSE state is delayed until the third progress variable reaches unity.  Our
intention is to allow a clear structure of the flame at low densities where
the stages will separate significantly, and where post-flame NSE evolution is
expected to be of relatively minor importance.  At high densities the NSE
timescale is short enough that the post-flame evolution of the NSE state
described below in section \ref{sec:postflame} takes over within a few grid
cells of the ADR flame stage ($\phi_1$).
Note that because $\phi_2$ and $\phi_3$ are not ADR variables, the
flame is not significantly thickened by the addition of multiple
burning stages.

By inspecting the evolution shown by the isochoric self-heating
calculations in section \ref{sec:flame_propagation}, we find that the
abundance evolution can be approximated by selected abundant elements in the
intermediate stages.  The stages then correspond to the following
transitions: stage 1, \carbon\ $\rightarrow$ \magnesium; stage 2,
\magnesium\ \& \oxygen\ $\rightarrow$ \silicon; stage 3, \silicon\ $\rightarrow$
NSE.  This allows the abundances to be related to the progress
variables, giving
\begin{eqnarray}
X_{^{12}\rm C} &=& (1-\phi_1)X_{\rm C}^0,\\
X_{^{16}\rm O} &=& (1-\phi_2)(1-X_{\rm C}^0),\\
X_{^{24}\rm Mg} &=& (\phi_1-\phi_2)X_{\rm C}^0,\\
X_{^{28}\rm Si} &=& (\phi_2-\phi_3),\\
\phi_3 &=& \sum X_{i,\rm NSE},
\end{eqnarray}
where $X_{\rm C}^0$ is the initial abundance of \carbon\ and the $X_{i,\rm
NSE}$ consists of the elements that make up the NSE abundances.
The relationship between the progress variables and the abundances, and their
evolution, is demonstrated in Fig. \ref{fig:3stage}.
The distribution of material among the NSE elements which together make up
$\phi_3$ is found from
\begin{equation}
\dot X_i = \dot\phi_3X_{i,\rm NSE}^{\rm sh}(\rho)
\end{equation}
and advecting the results in the usual way.  Here the $X_{i,\rm NSE}^{\rm sh}(\rho)$
result from the same self-heating calculation used to find $\Delta
Q_{NSE}$, described in the following section.  This also provides a
reasonable crossover to the post-flame evolution of the NSE element
abundances described below.  NSE material with approximately the correct
density dependent binding energy is monotonically produced by the flame.

\section{Input to the Flame model from network and NSE calculations }
\label{sec:network}

The three stage model flame front (Fig.~\ref{fig:3stage}) requires
{\em a priori} information characterizing the burn. In this section, we discuss 
the calibration of the energetics and timescales of the flame model. 
Using self-heating calculations allows us to make progress by removing the constraint 
on the size of time steps that occurs with an explicit hydrodynamics method (described
above).  Our principal approach for determining energetics and 
timescales is to perform self-heating simulations with the large nuclear 
reaction network, and this method works for most of the densities
of interest.  These self-heating calculations must still resolve the temporal 
evolution of the burn, which for the network calculations is accomplished
by employing a variable time step ordinary differential equation integrator. 
Although requiring small timesteps during some of the evolution, the calculations
are tractable.  At the lowest densities, however, the relatively long burning timescale 
forces us to obtain the abundances and binding energy of the NSE state by applying 
direct NSE calculations to the results of the self-heating network
calculations. These calculations are presented in this section,
and the direct NSE method is described in Appendix~\ref{sec:coulomb_correction}.
As mentioned above, the calibration presented here applies to the case of a
1:1 mass ratio of C and O. The model flame may be applied to any constant
mass ratio mixture, but would require the calculations described in this 
section to be repeated for the particular case (and use of the appropriate
input flame speed). The effect would be a different calibration, with longer or 
shorter timescales and a different energy release.  

The network we use contains 200 nuclides up to Kr, shown in
Fig.~\ref{fig:nuclides}.
We solve the evolution of the species and heat equation in time,
\begin{eqnarray}
\label{network.e}
\frac{dY_i}{dt} & = & \frac{1}{\rho N_{\rm A}} (-r_{\rm dest}^{(i)} +r_{\rm prod}^{(i)}) \ , \\
\label{heat.e}
\frac{dE}{dt} & = & \epsilon_{\rm nucl} + \frac{P}{\rho^2}\frac{d\rho}{dt} \ ,
\end{eqnarray}
where $Y_i$ is the abundance of species $i$ in mol g$^{-1}$, and E is the
internal energy per gram.  The term $r_{\rm dest}^{(i)} (r_{\rm prod}^{(i)})$
is the destruction (production) rate per volume of species $i$ owing to thermonuclear
reactions, which is in fact a sum over all the reaction paths in which the
nuclide participates.  $\epsilon_{\rm nucl}$ is the energy generation rate
due to thermonuclear burning, a sum of the reaction $Q$-values for each
pathway times their rates per gram of material.
We solve for the evolution of the nuclide abundances and
heat equation, eq.~(\ref{network.e}) and eq.~(\ref{heat.e}), using an
operator-split formalism.
For constant pressure, eq.~(\ref{heat.e}) can be written as
\begin{equation}
\frac{dT}{dt} = \epsilon_{\rm nucl} \cdot
\left[C_{\rm V} +
\left(\frac{P}{\rho^2} - \frac{\partial E}{\partial \rho}\right)
\frac{\partial P/\partial T}{\partial P/\partial \rho}\right]^{-1} \ ,
\label{temp.e}
\end{equation}
where all partial derivatives are either at constant $T$ or $\rho$, and
$C_{\rm V} = \frac{\partial E}{\partial T}$ is specific heat in constant density.
In writing eq.~(\ref{temp.e}),
we neglect the terms of change in internal energy due to the change in the species.
These terms are quite small compared with $\epsilon_{\rm nucl}$ except right at NSE.
Eq. (\ref{network.e}) is
stepped forward in time at a constant $T$ and $\rho$ using a semi-implicit
scheme combined with a sparse matrix solver \citep[see][and references
therein]{timmes+99}.  $T$ is updated by finite differencing (\ref{temp.e})
and solving explicitly. When $d\rho/dt=0$ (isochoric), eq.~(\ref{heat.e}) is
simplified to be $dE/dt=\epsilon_{\rm nucl}$, i.e., all
the energy input from nuclear burning is used to increase the internal energy
and hence the temperature. In this case, we update $T$ from the updated $E$
via nuclear burning by calling the equation of state, which takes into account the contribution
due to the change in the species. For constant $P$ (isobaric), some part of the
burning energy is used to expand the material. Using the constant P and
updated T (eq.~{\ref{temp.e}), $\rho$ can be updated by calling the equation of state.

The thermonuclear reaction rates are taken from an expanded 
version~\citetext{Schatz 2005, private communication} of the rate 
compilation REACLIB~\citep{thielemann+86,rauscher+00}.
We have also included the temperature-dependent nuclear partition functions 
provided by~\citet{rauscher+00}, both in the determination of 
the rates of inverse reactions and in our determination of 
NSE abundance patterns. To ensure that the rates obey detailed balance in NSE, 
the inverse reaction 
rates are derived directly from the forward rates based upon equilibrium 
equations that account for Coulomb interactions on the chemical 
potential (see
Appendix~\ref{sec:coulomb_correction}).\footnote{Common practice is to use
the reverse rates included in the REACLIB table, rather than calculating them
explicitly from the forward rates.  Note, however, that inverse reactions
for \carbon\ and \oxygen\ burning are currently omitted from REACLIB.} 
Our calculations demonstrate that the average binding energy and the abundances 
of the final NSE state from the self-heating calculation agrees with the results of
direct NSE calculations to within $1\%$.
We incorporate the effects of electron screening of thermonuclear reaction 
rates, adopting the relations for weak screening and strong screening 
provided by~\citet{wallace+82}. Finally, contributions from weak reactions 
are included using the rates provided by \citet{langanke+00,langanke+01}. 

\begin{figure}
  \includegraphics[width=\hsize]{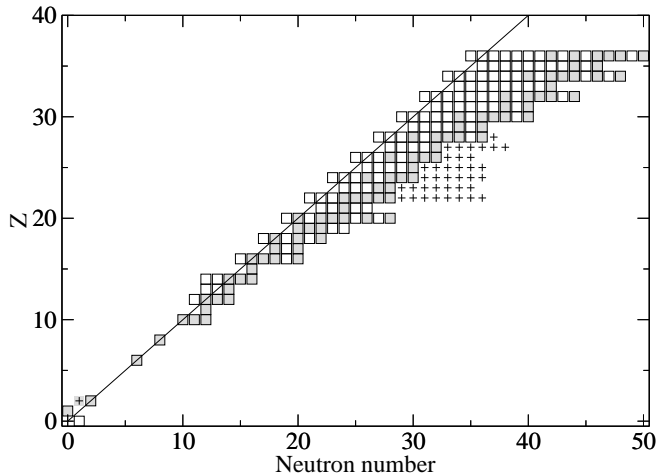}
\caption{Set of nuclides included in the 200 nuclide network (boxes) and
additional neutron rich nuclides added for the NSE calculations (crosses).
Shaded boxes indicate stable nuclides.
\label{fig:nuclides}}
\end{figure}

We apply this network in a collection of representative self-heating
calculations, which are intended to inform our flame model both of
the timescales required to reach NSQE or NSE and of the accompanying
nuclear energy release.
As discussed in \S~\ref{sec:flame_propagation}, an isochoric calculation
provides a good representation of the burning in an actual flame.  We also
use isochoric calculations below to discuss the importance of screening.
The initial temperature
is set as $T_9=1.7$, which is sufficient for \carbon\ and \oxygen\
ignition but lower than the final temperatures obtained at NSE for the
densities of interest. As with all calculations featured in this
work, the initial composition was taken to be half
\carbon\ and half \oxygen\ by mass.  The calculations were continued
until NSE was achieved, except at lower densities ($\rho < 10^7\gcc$),
for which the timescales to reach NSE are extremely long. Although the
self-heating calculations at very low densities do not yield the correct
NSE composition and released energy, they do provide temperatures close
to those of the NSE states. Hence, we apply the temperature achieved from
the self-heating calculation to the 
direct NSE calculation to get the 
correct released energy and composition.

The results of these systematic studies are presented in Table~\ref{tab:big}
and \ref{tab:nse_isobaric}, where, for the critical range of densities, we
tabulate the temperature achieved, the energy release in erg/g, and the mass
fractions of $^4$He and protons. The results from two other networks,
\textit{aprox19} and \textit{torch47}~\citep{timmes+99}, are listed as a
comparison for the isochoric case. As a reference point, the conversion of
an initial composition of half \carbon\ and half \oxygen\ by mass (with a
mean binding energy per nucleon of 7.829 MeV/nucleon) to pure \nickel\ (8.643
MeV/nucleon) yields 0.814 MeV/nucleon or equivalently $7.85 \times 10^{17}$
erg/g. The lower values of the energy release contained in
Table~\ref{tab:big} reflect the presence of significant concentrations of
$^4$He (7.074 MeV/nucleon) and protons; these are characteristic of NSE
distributions at higher temperatures, and can be seen to increase with
increasing density and temperature. Reliable determinations of the
compositions under these conditions are clearly crucial to our obtaining a
proper measure of the energy release, as described in our discussion of the
implementation of our flame energetics.

\begin{figure}
  \centering
  \includegraphics[width=\hsize]{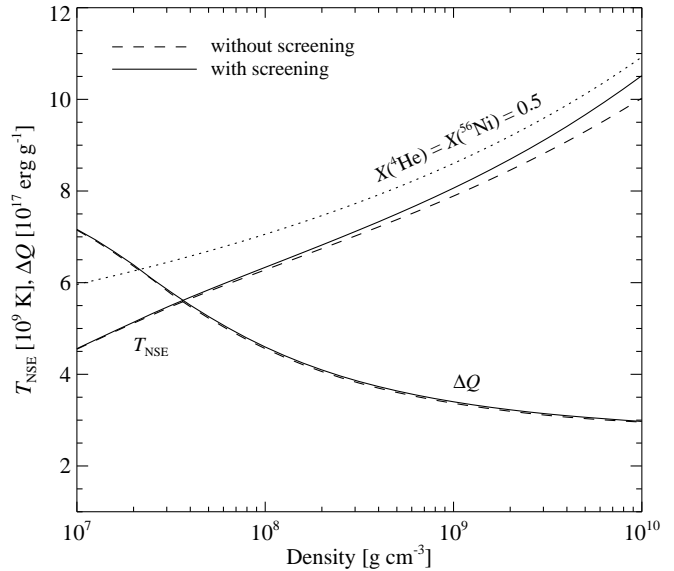}
  \caption{Energy released and NSE temperature for burning C/O as functions of density
   of the fuel for isochoric self-heating calculations with (solid lines) and without (dashed lines) 
   electron screening and Coulomb corrections. For comparison, the contour 
   corresponding to the condition (for $Y_e$ = 0.5) that the nuclear statistical
   equilibrium consists of equal parts by mass of $^4$He and $^{56}$Ni is shown (dotted line).  } 
  \label{fig:screen_energy}
\end{figure}
\begin{figure}
  \includegraphics[width=\hsize]{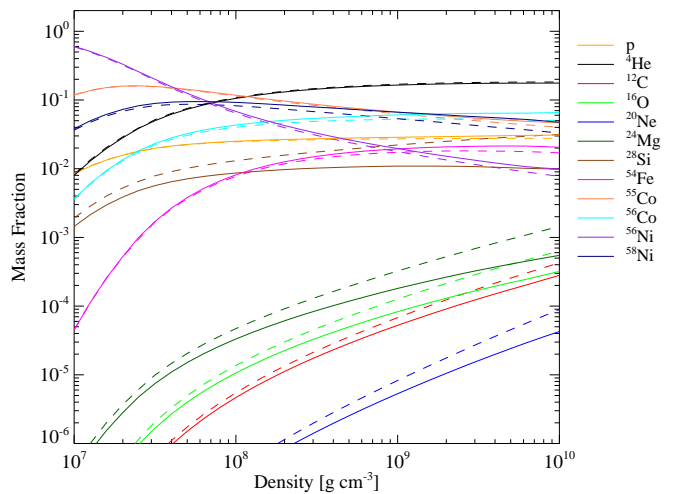}
\caption{Final (NSE) abundances of select species after C/O burning for the
cases shown in Fig.~\ref{fig:screen_energy}.  Dashed and solid lines have
the same meaning as there. }
\label{fig:compscreenpart}
\end{figure}

The ``ignition line", defined as the temperature-density relation of the
final NSE state from the self-heating calculations, is presented in
Fig.~\ref{fig:screen_energy}.  This is shown for the isochoric case in
order to facilitate the discussion of Coulomb screening corrections below.
At all of the densities shown the electrons are fully degenerate.
For comparison, we sketch the 
contour in the log$\rho-T_9$ plane corresponding to the condition (for 
$Y_e$ = 0.5) that the nuclear statistical equilibrium consists of equal 
parts by mass of $^4$He and $^{56}$Ni. The greatest proximity of our 
ignition line to the equal mass $^4$He/$^{56}$Ni contour is seen to occur at the 
highest densities (e.g. $10^9~{\rm g\,cm^{-3}}$) and correspondingly high 
temperatures. The free particle densities realized for the case 
$\rho$ = 10$^7$ are considerably lower, the matter is dominated by iron-peak 
nuclei, and the corresponding mean binding energy per nucleon most closely 
approaches that of pure $^{56}$Ni. Note that our ignition line can 
approach - but never overlay - the equal mass $^4$He/$^{56}$Ni contour for a very 
straightforward reason. Recall that our initial condition consisted of 
equal masses of $^{12}$C and $^{16}$O; the mean binding energy per nucleon 
for this mixture is $\sim$ 7.8 MeV/nucleon. However, the mean B/A for a mixture 
of equal parts by mass $^4$He and 
$^{56}$Ni is also B/A$\sim$7.8 MeV/nucleon. It follows 
that the $^4$He-$^{56}$Ni contour is physically not achievable for this
fuel composition.

\subsection {Coulomb Correction and Energetics}
\label{sec:coulomb}

The impact of our inclusion of Coulomb corrections and screening effects both
on the NSE temperatures achieved and on the energetics are illustrated in
Fig.~\ref{fig:screen_energy}. A detailed discussion of the manner in which
Coulomb and screening effects are utilized in the determination of screened
reaction rates and nuclear statistical equilibrium configurations is
presented in Appendix~\ref{sec:coulomb_correction}.  In order to facilitate
comparison, the case without screening neglects both reaction rate screening
and Coulomb corrections to the equation of state (EOS).  This necessitates an
isochoric comparison due to the lower pressure with Coulomb corrections in
the EOS.  We note from the figure that even at the highest densities, where
screening effects are most pronounced ($\Gamma_{^{56}\rm Ni}=Z^{5/3}e^2(4\pi
n_e/3)^{1/3}/kT = 10$), the differences in energy release is negligible and
the differences in temperatures achieved are quite small. For the limiting
case of $\rho=10^{10}$ g cm$^{-3}$, the temperature achieved without
screening and Coulomb correction is $T_9 = 10$ while with screening it is
$T_9=10.5$, increasing by $5\%$. For the same conditions, the difference in
energy release is less than $1\%$.

In order to understand how plasma Coulomb corrections affect the final
temperature and energy release, it is useful to look at how the final (NSE)
composition changes in detail.  
Fig.~\ref{fig:compscreenpart} demonstrates the difference of the NSE
compositions with screened (\emph{solid} lines) and unscreened (\emph{dashed}
lines) network calculations.  The difference is again quite modest.  Generally
the inclusion of plasma Coulomb corrections energetically favors assembling
fewer, higher charge ions from lower charge ones.  Consistent with this, we
find that the abundances of the high-charge species increase at the expense
of low-charge ones.  In light of the minor change in abundances, the
ideal-gas contribution to the heat capacity (which is related to the mean
molecular weight of the ion gas) changes very little. Thus the the difference
in temperature at the high density between the screened and unscreened cases
is likely due to the difference in the EOS being used.  The Coulomb
corrections decrease the heat capacity of the ion gas, allowing higher
temperatures to be reached for the same energy release.

The degree of agreement between the energy release with and without screening
terms is remarkable.  A clue to the origin of this match is that there is one
exception to the general rule that Coulomb enhances heavy element abundances:
protons.  Although the $^4$He, $^{12}$C, and $^{16}$O abundances are decreased
in the screened case, protons actually increase.  This is because we are
working with the constraint that $Y_e=0.5$, while $^{56}$Ni is by far not
the most abundant Fe-group nucleus, with more neutron-rich nuclides like
$^{56}$Co and $^{58}$Ni being favored.  Thus it appears that although
high-charge nuclei are favored with screening included, this is offset in
energy release by the concomitant increase in proton abundance.

\begin{deluxetable*}{cccccc}
\tabletypesize{\scriptsize}
\tablewidth{0pt}
\tablecaption{NSQE and NSE timescales from self-heating network calculations with the 200 nuclide network. 
}
\tablehead{ 
\multicolumn{1}{c}{  } & \colhead{ } &  \colhead{max. \silicon\ abund.} & \colhead{90\% of \nickel\ abund.} &  \colhead{90\% of Fe-peak nucleus} &  \colhead{90\% of NSE nuclei} \\ 
\colhead{log($\rho$)} &  \colhead{$T_{\rm NSE}$} & \colhead{$\tau_{\rm NSQE}$} & \colhead{$\tau_{\rm NSE}$} &  \colhead{$\tau_{\rm NSE}$} &  \colhead{$\tau_{\rm NSE}$} \\ 
\colhead{($\gcc$)} & \colhead{({$10^9$} K)} &  \colhead{(s)} & \colhead{(s)} & \colhead{(s)}  & \colhead{(s)}   
} 
\startdata
7.0  & 4.56 & 2.21E-03 & 2.90E+00 & 2.90E+00 & 2.86E+00\\  
7.1  & 4.75 & 5.46E-04 & 6.62E-01 & 6.62E-01 & 6.65E-01\\
7.2  & 4.95 & 1.39E-04 & 1.70E-01 & 1.70E-01 & 1.68E-01\\
7.3  & 5.13 & 3.75E-05 & 4.94E-02 & 4.94E-02 & 4.56E-02\\
7.4  & 5.32 & 1.07E-05 & 1.19E-02 & 1.19E-02 & 9.40E-03\\
7.5  & 5.50 & 3.28E-06 & 4.33E-03 & 7.16E-04 & 3.02E-03\\
7.6  & 5.68 & 1.08E-06 & 1.72E-03 & 2.26E-04 & 1.02E-03\\
7.7  & 5.85 & 3.81E-07 & 7.80E-04 & 8.08E-05 & 3.85E-04\\
7.8  & 6.01 & 1.45E-07 & 3.61E-04 & 3.22E-05 & 1.46E-04\\
7.9  & 6.17 & 5.85E-08 & 1.68E-04 & 1.45E-05 & 5.64E-05\\
8.0  & 6.33 & 2.50E-08 & 8.01E-05 & 7.08E-06 & 2.27E-05\\
8.1  & 6.50 & 1.12E-08 & 3.87E-05 & 3.66E-06 & 9.56E-06\\
8.2  & 6.66 & 5.19E-09 & 1.89E-05 & 1.97E-06 & 4.22E-06\\
8.3  & 6.82 & 2.50E-09 & 9.41E-06 & 1.10E-06 & 1.94E-06\\
8.4  & 6.99 & 1.26E-09 & 4.77E-06 & 6.01E-07 & 9.35E-07\\
8.5  & 7.15 & 6.58E-10 & 2.48E-06 & 3.45E-07 & 4.67E-07\\
8.6  & 7.33 & 3.58E-10 & 1.30E-06 & 2.03E-07 & 2.41E-07\\
8.7  & 7.50 & 2.02E-10 & 7.00E-07 & 1.20E-07 & 1.28E-07\\
8.8  & 7.68 & 1.19E-10 & 3.82E-07 & 7.17E-08 & 7.01E-08\\
8.9  & 7.87 & 7.13E-11 & 2.12E-07 & 4.32E-08 & 3.91E-08\\
9.0  & 8.06 & 4.42E-11 & 1.21E-07 & 2.52E-08 & 2.24E-08\\
9.1  & 8.26 & 2.80E-11 & 6.86E-08 & 1.55E-08 & 1.30E-08\\
9.2  & 8.47 & 1.81E-11 & 3.97E-08 & 9.50E-09 & 7.64E-09\\
9.3  & 8.69 & 1.19E-11 & 2.32E-08 & 5.88E-09 & 4.58E-09\\
9.4  & 8.92 & 7.86E-12 & 1.37E-08 & 3.61E-09 & 2.78E-09\\
9.5  & 9.16 & 5.24E-12 & 8.22E-09 & 2.23E-09 & 1.71E-09\\
9.6  & 9.40 & 3.51E-12 & 4.98E-09 & 1.41E-09 & 1.07E-09\\
9.7  & 9.60 & 2.38E-12 & 3.07E-09 & 8.98E-10 & 6.78E-10\\
9.8  & 9.94 & 1.62E-12 & 1.91E-09 & 5.90E-10 & 4.32E-10\\
9.9  & 10.2 & 1.10E-12 & 1.19E-09 & 3.86E-10 & 2.78E-10\\
10.0 & 10.5 & 7.45E-13 & 7.66E-10 & 2.49E-10 & 1.81E-10\\
\enddata
\label{tab:nse}
\tablecomments{The NSQE timescale is defined as the elapsed time in a simulation between
the point at which 90\% of the C has burned to the maximum of \silicon\ abundance. The time scales to NSE were determined as the elapsed time from the maximum abundance of $^{28}$Si, to the 90\% of the maximum $^{56}$Ni abundance, to the 90\% of the maximum abundance of the most abundant Fe-peak nucleus, and to 90\% of the maximum number of NSE nuclei, respectively. NSE nuclei were $^4$He and isotopes of Cr, Fe, Co, Ni, and Zn.}
\end{deluxetable*}

\subsection {NSQE and NSE timescale calculations}
\label{sec:timescales}

\begin{figure}
  \includegraphics[width=\hsize]{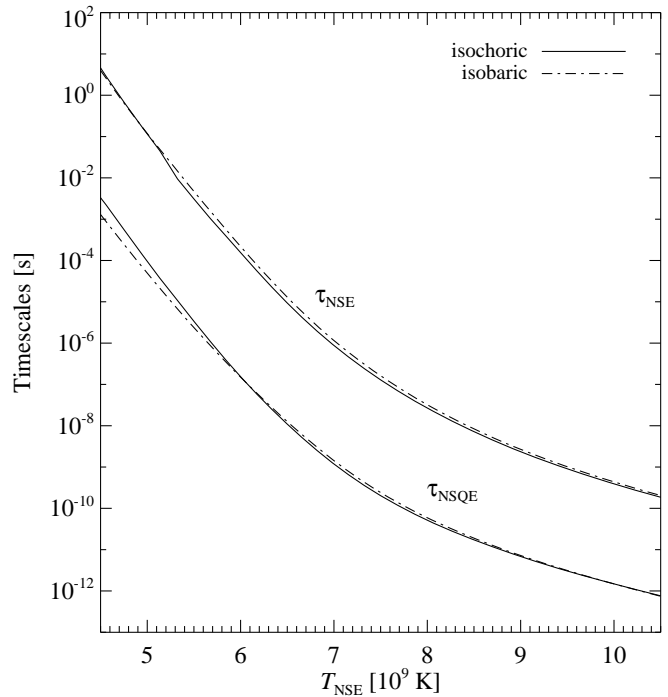}
\caption{Nuclear burning timescales versus ash temperature of C/O fuel calculated
     by 200 nuclide self-heating network.
     Shown are curves for the timescales to NSQE and NSE for isochoric (\emph{solid lines}) and for  
     isobaric (\emph{dashed lines}) one-zone calculations.}
\label{fig:time_T}
\end{figure} 

\begin{figure}
  \includegraphics[width=\hsize]{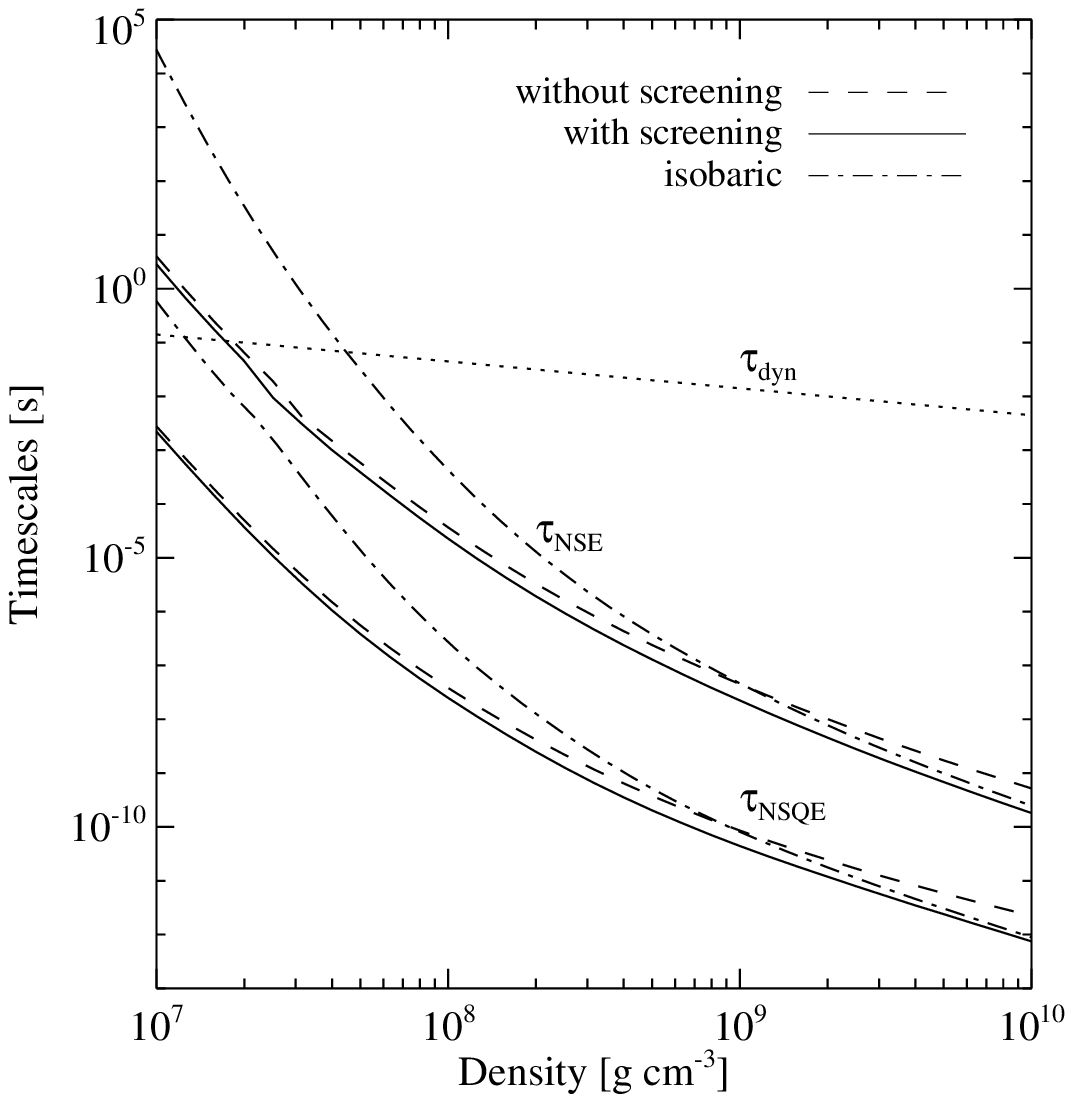}
\caption{Nuclear burning timescales versus density of C/O fuel calculated
     by 200 nuclide self-heating network.
     Shown are curves for the timescales to reach NSQE and NSE with (\emph{solid lines}) or without
     (\emph{dashed lines}) screening and Coulomb corrections for an isochoric
     burn and with corrections for an isobaric burn (\emph{dash-dot lines}). Superposed is the dynamical timescale (\emph{dotted line}), $\tau_{\rm dyn}$.}
\label{fig:screen_time}
\end{figure} 

As noted above, the three-stage flame model requires timescales for the second
and third stages of burning. These timescales must be determined in advance
of simulations with the three-stage flame model, and accordingly we investigated
determining these timescales with the self-heating networks available to us,
and the effect of screening on these timescales.

The timescale to NSQE was defined as the elapsed time in a simulation between
the point at which 90\% of the C has burned and the maximum abundance of $^{28}$Si.
Defining the timescale to NSE requires a little more care, and we investigated
several approaches. In all cases, the timescale was measured as the elapsed
time from the maximum abundance of $^{28}$Si. The results of these studies are
shown in Table~\ref{tab:nse}. In Table~\ref{tab:nse} we list the NSE 
timescales defined in several ways: the time when 90\% of the maximum \nickel\ abundance is 
reached; the time when 90\% of the maximum abundance of the most abundant Fe-peak isotope is reached; the time when 90\% of the maximum abundance of NSE nuclei is reached. NSE nuclei include \helium\ and isotopes of  Cr, Fe, Co, Ni, and Zn. At the lowest density end, Ni dominates the NSE compositions, so all the approaches yield the same NSE timescales. At the highest density end, the dominant NSE composition is a mixture of He and Fe-peak isotopes. By including Fe-peak nuclides other than Ni, we can see that the timescales differ at most within a factor of 3 at the high density end. Including He decreases the timescale further by a factor of 1.5 at the high density end.  In this paper, we choose the last definition as NSE timescale in implementing the flame model. 
The NSQE and NSE timescales from the self-heating calculations are fitted by
\begin{equation}
\label{eq:tnsqe}
\tau_{\rm NSQE} = \exp{(182.06/T_9 - 46.054)} 
\end{equation}
and
\begin{equation}
\label{eq:tnse}
\tau_{\rm NSE} = \exp{(196.02/T_9 - 41.645)} \ . 
\end{equation}
We express the timescales as functions of final temperature,
which is determined by the final
density and energy release in self-heating calculations.
This is a quite robust parameterization as seen from the comparison of
timescales determined from isochoric and isobaric calculations in
Fig.~\ref{fig:time_T}.
We note that our
NSQE/NSE timescales are consistently lower than those
of~\citet{khokhlov91+dd}, but we are unable to determine the reason for this
difference.
 
The influence of screening on the burning timescales, as inferred from our
self-heating calculations, is revealed in Fig.~\ref{fig:screen_time}.  Once
again we use an isochoric case to isolate the effects of rate screening.  In
contrast to our findings for temperature and energy, the impact of the
inclusion of Coulomb and screening effects is seen here to decrease the
timescales to achieve NSQE and NSE by factors of 2--3. This reflects the
level of individual screening enhancements for the critical reactions that
govern the flows into the iron-peak region.  An isobaric burn,
representative of what will occur in the star, is also shown, and has much
longer timescales at lower densities due to the lower temperature on
expansion.  For comparison, the dynamical timescale for homologous collapse,
$\tau_{\rm dyn} = (24 \pi G \rho)^{-1/2} = 0.14(10^{7}\;\gcc/\rho)^{1/2}~{\rm
s}$, is superposed (\emph{dotted line}) on the figure. Note that
$\tau_{\mathrm{NSE}} > \tau_{\mathrm{dyn}}$ for  $\rho < 4 \times 10^7\;\gcc$
and $\tau_{\mathrm{NSQE}} > \tau_{\mathrm{dyn}}$ for $\rho < 1.5\times
10^7\;\gcc$.

\section{Post-flame energy release and neutronization}
\label{sec:postflame}

Once the flame has passed through a piece of stellar material, the remaining
ashes have burned to NSE, which is predominantly comprised of iron-peak
nuclei, $^4$He, and free protons.  This NSE composition, however, should not be considered
static. It changes as the density decreases, either in a rising bubble or due to
the expansion of the star, resulting in a varying ionic heat capacity and
more importantly causing the difference in nuclear binding energy between the
initial \carbon/\oxygen\ fuel and the NSE state to vary during
the course of the hydrodynamic evolution of the supernova.  During this
extended evolution, there is time for weak interactions to take
place in the ash material.  Dominated by electron captures, these processes
lower $Y_e$ and emit neutrinos.  Below we discuss these two physically
distinct effects which bring about changes in the NSE abundances.  The method
used to calculate (Coulomb-corrected) NSE is described in Appendix
\ref{sec:coulomb_correction}, with the set of nuclides used in our network
calculation supplemented with neutron-rich Fe peak elements as shown in
Fig.~\ref{fig:nuclides}.

\subsection{NSE adjustment to the evolving state variables $\rho$ and T}
\label{sec:nseadjust}

The hydrodynamic evolution of the ashes, i.e. the expansion following the
burn, causes a decrease in $\rho$ and $T$ on a hydrodynamic timescale, which
in turn causes a change in the NSE mass fractions as the nucleons reorganize
to maximize entropy for the new thermodynamic state. As long as the
hydrodynamic timescale substantially exceeds the charged-particle (strong)
nuclear reaction timescale, the approximation of instantaneous readjustment
of the composition of the ashes to the NSE appropriate for the local
thermodynamic state is a good one.

\begin{figure}
\includegraphics[width=\hsize]{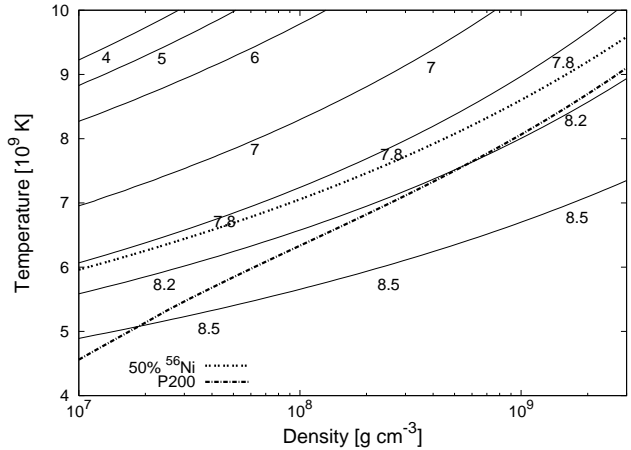}
\caption{Contours of binding energy per nucleon, $q$, in MeV per nucleon, for
the NSE state with $Y_e=0.5$.  Also shown is as a function of density the final temperature 
reached (dot-dashed line) by the 200 nuclide self-heating network calculation starting from an 
equal mixture of C and O by mass and burning to NSE.  The line
of 50\% \nickel\ mass fraction (dotted line) in a two-component NSE (\helium\ and
\nickel) indicates the temperature above which light particles dominate.
\label{fig:qcontour0.5}
}
\end{figure}

The electron fraction, defined as $Y_e = \sum_i {Z_i \frac{X_i}{m_i}}/ \sum_i
{A_i \frac{X_i}{m_i}}$, where $m_i$ is the nuclear mass of species $i$, of the \carbon/\oxygen\ mixture is 0.5.
Weak interactions are assumed to be negligible during the burning phase of
the flame since the weak interaction timescale far exceeds the burning
timescale. At first, the composition of the hot, unexpanded ashes has a
significant fraction of the mass in free protons (0 MeV) and \helium\
(7.074 MeV/nucleon).  For example at $\rho_8 = 1, \mathrm{X}_\mathrm{p}
= 0.024, \mathrm{X}_{\mathrm{He}} = 0.095$ and at $\rho_8 = 10,
\mathrm{X}_\mathrm{p} = 0.027, \mathrm{X}_{\mathrm{He}} = 0.135$.
The remainder predominantly consists of a mix of strongly bound Fe-peak
nuclei, dominated by \iron.  The nuclear binding energies
reached in the hot ashes immediately following the burn are therefore
significantly smaller than the binding energies obtained by a cold
NSE dominated by \nickel\  and lie along the dot-dashed line in
Fig.~\ref{fig:qcontour0.5}.


The binding energy gain per nucleon of such a mixture compared to
the initial \carbon/\oxygen\ fuel, which has a binding energy per
nucleon of 7.828 MeV, varies with density but remains for $\rho_8 >
1$ in the vicinity of 0.4 MeV.  As the material expands and cools, NSE
progressively favors heavier nuclei.  The reason for the shift from an
alpha rich NSE to an alpha poor NSE during an expansion is that lower
entropies allow for a reduction of the total number of particles, since
the number of accessible microstates of the ensemble has decreased.
The light particles are reassembled into fewer but more tightly bound
Fe-peak nuclei. At constant $Y_e=0.5$, the composition would reach
virtually 100\% \nickel\ at freezeout temperatures. In other words, at
constant $Y_e = 0.5$, the binding energy would shift from $\sim$ 8.2
to 8.643 MeV per nucleon and an additional $\sim$ 0.4 MeV of binding energy would be
released per nucleon during the expansion phase. This is a significant
fraction of the total energy released and must be accounted for.

Furthermore, because the average nucleon number per nucleus $\bar{A}=
\sum_i{A_i\frac{X_i}{m_i}}/ \sum_i{\frac{X_i}{m_i}}$ grows as the NSE
distribution is further shifted with decreasing temperature toward
the iron-peak nuclei, the ionic heat capacity decreases. Because we have
conditions that are at the borderline of electron degeneracy, we estimate
this to have a small effect.

\subsection{Neutronization}
\label{sec:neutronization}

Virtually all extant one-dimensional calculations of SNe Ia explosions find that the innermost regions (the inner $\approx$ 0.2 M$_\odot$) suffer
sufficient changes in $Y_e$, due to electron captures, to significantly change the compositions \citep{nomoto+84,khokhlov91+pd,iwamoto+99}.
The use of recent electron capture rates, based on shell model diagonalization, reduces the neutron excess in the central regions somewhat \citep{brachwitz+00},
the gross features however remain. Neutronization is a consequence of the high Fermi energies and 
the correspondingly high rates of electron capture reactions under these conditions. 
The time rate of change of the electron fraction $\dot{Y_e}$, also referred
to as the neutronization rate, and the amount of energy carried away by
neutrinos per gram of stellar material per second $\dot{E^{\nu}}$ are
calculated as NSE averages by convolving tabulated weak interaction
rates~\citep{langanke+01} with the NSE abundances. We consider electron
decay, positron captures, electron captures and positron decay of pf-Shell
nuclei as well as free nucleons, so that
\begin{eqnarray}
\dot{Y_e}&=&\sum_i{\frac{X_i m_u}{m_i} \Big(\lambda^{ed}_{i}+\lambda^{pc}_i-\lambda^{ec}_i-\lambda^{pd}_i \Big)}
\\
\dot{E^{\nu}}&=&\sum_i{\frac{X_i m_u}{m_i} \Big( \lambda^{\bar{\nu}}_{i}+\lambda^{\nu}_{i}\Big)}
\end{eqnarray}
where we are following the notation from \citet{langanke+01} in which $\lambda^{ed}, \lambda^{pc},\lambda^{ec}, \lambda^{pd}$ 
are given in units of $\mathrm{s}^{-1}$ and $\lambda^{\bar{\nu}}, \lambda^{\nu}$ are given in units of $\mathrm{MeV} \; \mathrm{s}^{-1}$. 
Neutrinos emitted are taken to be free-streaming. 

At high densities ($\rho_9 \gtrsim 1$) the NSE averaged neutronization rate for self 
conjugate matter is rather fast with $-0.05\ {\rm s^{-1}} \gtrsim \dot Y_e
\gtrsim -0.5\ {\rm s^{-1}}$ (see Fig. \ref{fig:yecontour0.5}). The rate of neutronization is strongly dependent on the electron fraction 
(compare Figs. \ref{fig:yecontour0.5} and \ref{fig:yecontour0.48}),
since electron capture rates are overall decreasing with increasing neutron number for a given element.  
Due to the strong decrease of the rate of neutronization with decreasing electron fraction, one typically finds the composition of these inner regions to consist of moderately neutron-rich, strongly bound iron-peak nuclei, viz; $^{56}$Fe, $^{54}$Fe, and $^{58}$Ni.

\begin{figure}
\includegraphics[width=\hsize]{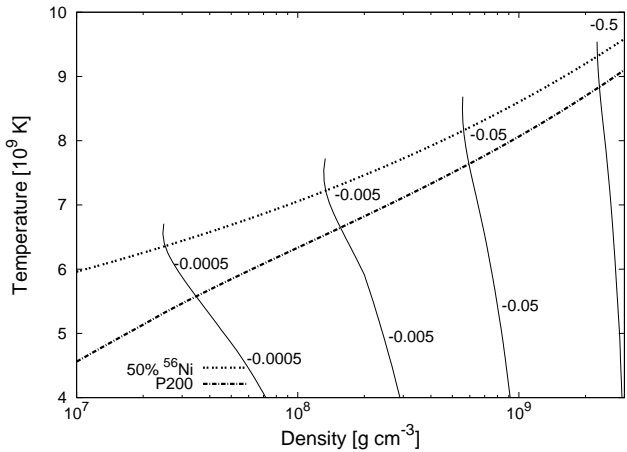}
\caption{Contours of time derivative of electron fraction, $\dot Y_e$, due to
electron and positron capture and emission, evaluated at $Y_e=0.5$.  Other lines
are as in Fig. \ref{fig:qcontour0.5}.
}
\label{fig:yecontour0.5}
\end{figure}

\begin{figure}
\includegraphics[width=\hsize]{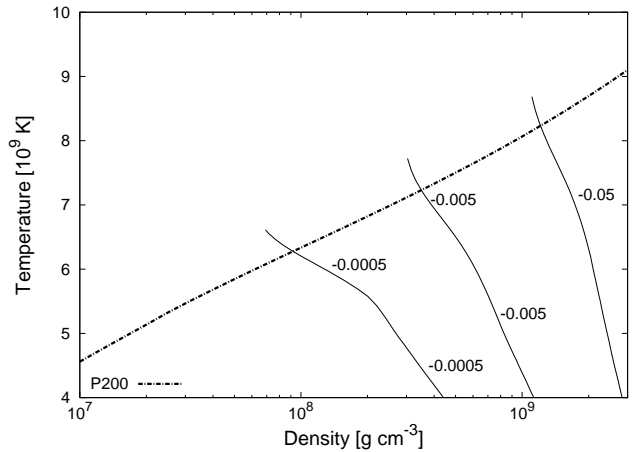}
\caption{Contours of time derivative of electron fraction, $\dot Y_e$, due to
electron and positron capture and emission, evaluated at $Y_e=0.48$.  Other lines
are as in Fig. \ref{fig:qcontour0.5}.
}
\label{fig:yecontour0.48}
\end{figure}

If neutronization occurs rapidly enough to lower $Y_e$ to $\sim 0.48$, then the binding energy per nucleon of the neutron rich NSE state compared to the NSE state with $Y_e = 0.5$
is increased by $\sim (0.2-0.3)$ MeV for a range of given temperatures and densities (compare Figs. \ref{fig:qcontour0.5} and \ref{fig:qcontour0.48}).
In the cold NSE limit, where $^{54}$Fe (8.736 MeV/nucleon) and $^{58}$Ni (8.732 MeV/nucleon) 
are the dominant nuclei of the $Y_e = 0.48$ NSE abundance distribution, the final binding energy per nucleon increase compared to NSE of self conjugate matter,
which is entirely dominated by $^{56}$Ni, is on the order of 0.14 MeV. 
Not all of this energy, however, remains available to expand the star. Neutrino losses accompanying neutronization impact the energetics: some fraction of the
energy release (being somewhat dependent upon the density and corresponding electron Fermi energy) will be lost in neutrinos (see Figs. \ref{fig:econtour0.5} \& \ref{fig:econtour0.48}).

\begin{figure}
\includegraphics[width=\hsize]{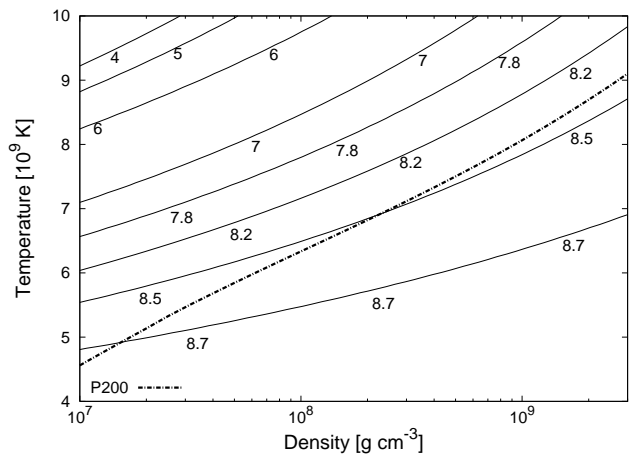}
\caption{Contours of binding energy per nucleon, $q$, in MeV per nucleon, for
the NSE state with $Y_e=0.48$.  Other lines
are as in Fig. \ref{fig:qcontour0.5}.
}
\label{fig:qcontour0.48}
\end{figure}


\begin{figure}
\includegraphics[width=\hsize]{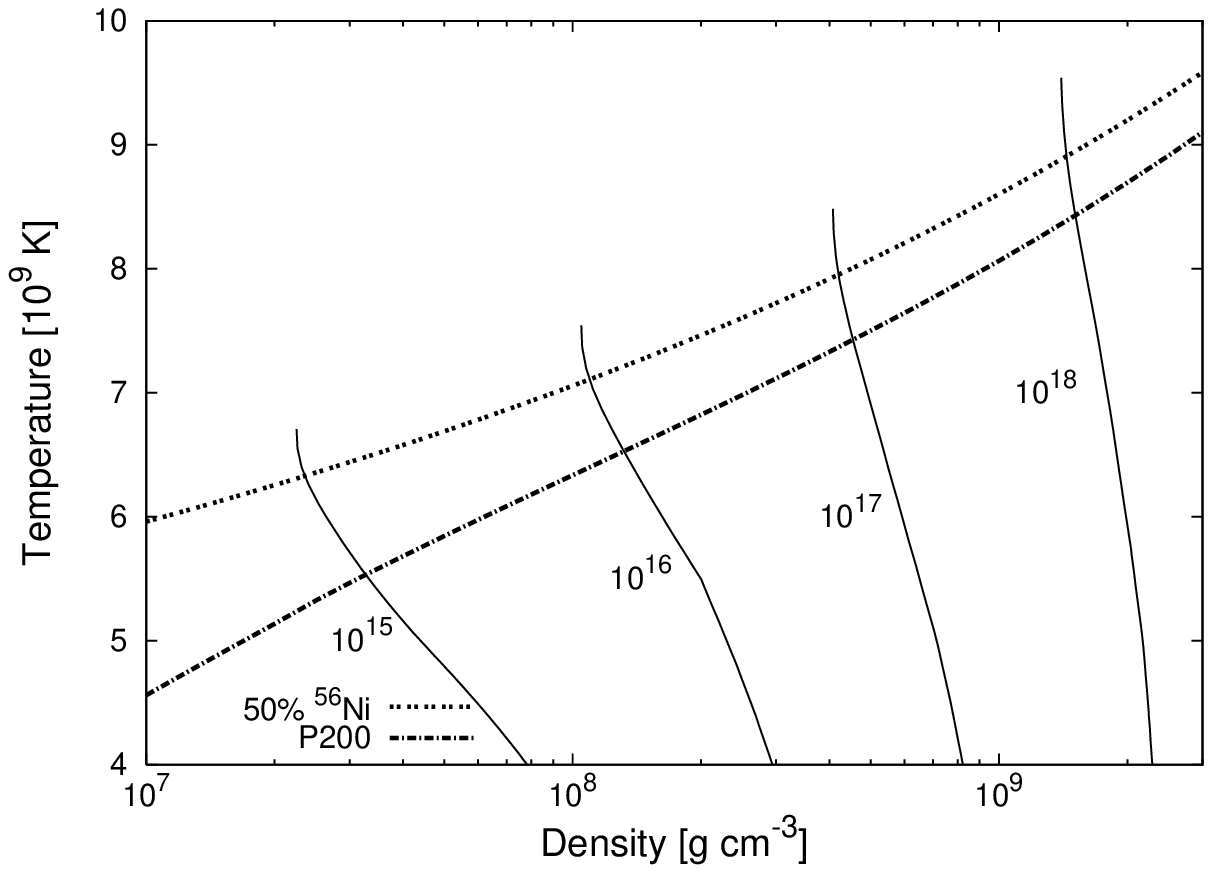}
\caption{Contours of neutrino energy loss rates, $\epsilon_\nu$, during
electron and positron captures and emissions, evaluated at $Y_e=0.5$.  Other
lines are as in Fig. \ref{fig:qcontour0.5}}.
\label{fig:econtour0.5}
\end{figure}

\begin{figure}
\includegraphics[width=\hsize]{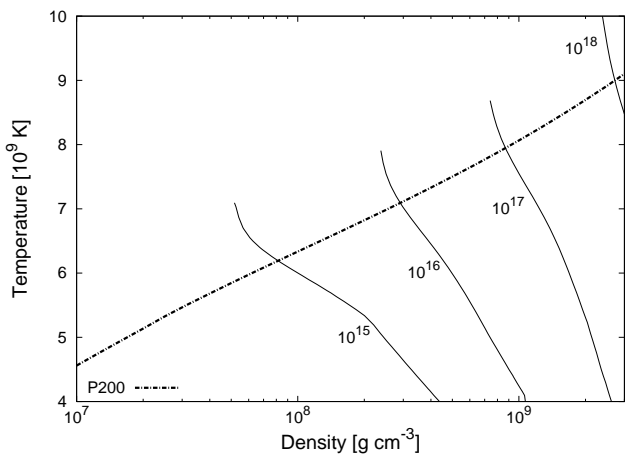}
\caption{Contours of neutrino energy loss rates, $\epsilon_\nu$, during
electron and positron captures and emissions, evaluated at $Y_e=0.48$.  Other
lines are as in Fig. \ref{fig:qcontour0.5}}.
\label{fig:econtour0.48}
\end{figure}

Another important effect of neutronization lies in its effect on the Fermi level of the electrons. As electrons get captured, $Y_e$ decreases and since relativistic
electron degeneracy pressure scales as $(\rho Y_e)^\frac{4}{3}$ this results in a change in the ratio of electron degeneracy pressure to ionic pressure. Because we are close to the borderline of degeneracy, it is difficult to estimate the impact of this effect on the thermal histories of tracer
particles being followed in the simulation.

The impact of neutronization on the rise times of hot bubbles and the subsequent growth of the Rayleigh Taylor instabilities could help to distinguish between competing ignition scenarios. In addition, it affects the relation between the composition of the progenitor white dwarf and the mass of \nickel\ synthesized in the explosion \citep{timmes.brown.ea:variations}.


\subsection{Reduced Set of Nuclei for Evolution in NSE}
\label{sec:nseimplementation}

While the NSE state is calculated using many nuclear species, it is not
feasible to store the abundance of each of these at every grid point.  In
reality only a small number of quantities which are derivable from the set of
abundances, $\{X_i\}$, are essential to accurate hydrodynamic and energetic
simulation of the NSE material.  These are the electron fraction, $Y_e$, the
average number of nucleons per ion, $\bar A$, the average binding
energy per nucleon, $\bar q = \bar Q/\bar A$, and the coulomb coupling
parameter $\Gamma = \overline{Z^{5/3}}e^2(4\pi n_e/3)^{1/3}/kT$, where $n_e$
is the electron number density.
The quantities $Y_e$, $\bar A$,
and $\Gamma$ are required to determine the local pressure from density and
internal energy content, while $\bar q$ provides a measure of the energy
stored as rest mass energy of (comparatively less bound) light nuclei, which
is released as the NSE state evolves.

Working in the approximation that $m_i = A_i$ a.m.u. reduces several
quantities to simple sums over the abundance set, $Y_e = \sum_iX_iZ_i/A_i$,
$1/\bar A = \sum_i X_i/A_i$, and $\bar q = \sum_i X_iQ_i/A_i$.  This
approximation is accurate at the level of $10^{-4}$ for these averages, but
is unsatisfactory for the actual calculation of NSE due to its dependence on
small relative differences between species.  With this simple linear form, we
see that it is possible to select a set of abundances $\{X'_i\}$ which give
the same $Y_e$, $\bar A$ and $\bar q$, and give the correct mixing evolution
of these quantities (essential for an Eulerian code), but which consists of
only a select few elements.  This representative set can then be used in the
simulation without losing information about the energetic evolution of
the material, giving appropriate Lagrangian temperature histories which can
be used for post-processing.

Choosing this reduced set is much like finding basis vectors which span the
3D space $(Y_e,1/\bar A, q)$, except that $\sum X'_i$ is constrained to be 1,
so that the set must both span and bound the available space.  We restrict
ourselves to elements which exist in nature and thus have known $(Z/A, 1/A,
Q/A)$.  The goal of this method is to utilize existing hydrodynamic tools
which are known to treat the $X_i$ properly.  An alternative is for the basic
quantities $(Y_e,1/\bar A, q)$ to be treated directly by the hydrodynamics
method, as is done by \citet{khokhlov95}.

The quantities $Y_e$, $\bar A$, and
$\bar q$ are three of the four quantities tracked by the nuclear kinetic
equations of \citet{khokhlov95} (and related references).  The fourth is a
variable to track the relaxation from NSQE to NSE, which they approximate to
have no effect on the energy release, and which corresponds to our third
flame stage.  \citet{khokhlov+01} makes the additional approximation that
$\bar A$ (their $Y_{\rm i}$) is constant in the ashes.

We do not treat $\Gamma$ in any special way.  It is expected, however, that
because the representative nuclei do not have very different $Z$ from the real
ones, the error is not too bad. NSE is generally a mixture of $^4$He and
many species near $^{56}$Ni, making $\Gamma$ fairly well matched if all the
other quantities are matched.  Thus the benefit of developing a 4D matching
method that includes $\overline{Z^{5/3}}$ does not appear necessary for the
Ia supernova problem, where $\Gamma$ is moderate ($\simeq 0.5$), and
declining, during the evolution.  The simple dependence on a $\Gamma$ which
contains $\overline{Z^{5/3}}$ is also not a general feature of Coulomb
corrections, but only of a particular approximation used for EOS corrections.

\begin{figure}
\includegraphics[width=\hsize]{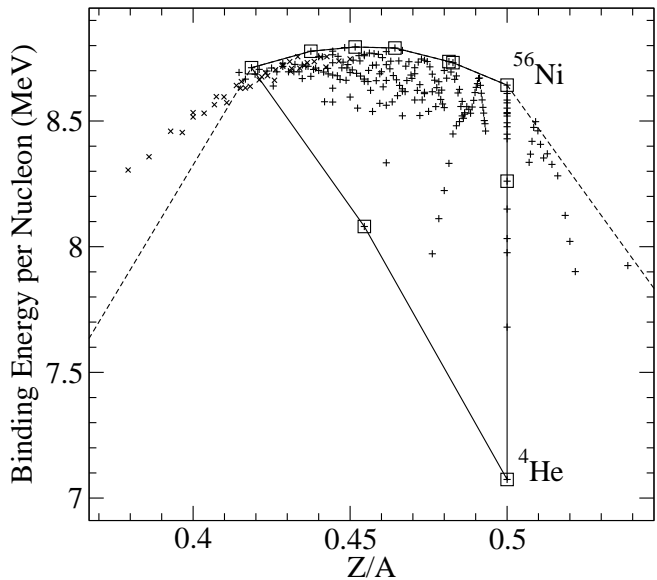}
\caption{
\label{fig:BYe}
Binding energy per nucleon $Q/A$ vs. $Z/A$ (respectively $\bar q$ and $Y_e$
of a pure sample) for nuclei used in simulations.  Shown are the 200 nuclei
in the dynamic network calculation ($+$), and those that were added to extend
the $Y_e$ range of the NSE calculation ($\times$).  Dashed lines extend
toward neutrons on the left and $^1$H on the right; $^3$He is also off the
plot.  Boxes indicate those included in the representative set.  The set of
heavy nuclei is chosen to allow spanning of the entire possible area of $q$
and $Y_e$ (outlined) and also allows accurate representation of $\bar A$,
precise matching of $\bar A$ is achieved in most cases in all but the lowest
$Y_e$ portion of the solid-outlined area.
}
\end{figure}
The set of elements we have chosen, which includes several necessary for
evolution of the flame, consists of neutrons, $^1$H, $^4$He, $^{22}$Ne,
$^{24}$Mg, $^{56}$Ni, $^{58}$Ni, $^{56}$Fe, $^{62}$Ni, $^{64}$Ni, and
$^{86}$Kr.  (The full set in the simulation also includes $^{12}$C,
$^{16}$O, and $^{28}$Si, used in the burning stages.)  These are shown in
Fig.~\ref{fig:BYe} along with the full set used in the nuclear network and
NSE calculation.  The set of neutron-rich heavy elements is used to follow in
detail the energy release as the material neutronizes.
At most
four species are used to construct a given $(Y_e,1/\bar A, \bar q)$, where
the species are chosen by comparing these coordinates to the face planes of
the four-sided polyhedra formed by four-element sets, starting from
($^4$He, $^{24}$Mg, $^{56}$Ni, $^{58}$Ni) and walking to more neutron rich
elements as necessary.  If the desired combination lies outside of polyhedra
available with these elements, $\bar A$ is not matched and the three species
which form the triangle enclosing the $Y_e$ and $\bar q$ of the state are
used.  The set above allows us to match $Y_e$ and $\bar q$ exactly to that of
the NSE state for $Y_e >0.42$.  $\bar A$ is also matched exactly in most
cases, except for $Y_e$ near 0.42, where it may vary from the NSE value by up
to 10\%.


\subsection{Energy Evolution in NSE}

The NSE properties of the material are used to evolve the mass-specific
thermal energy $\mathcal E$, much like a nuclear network would evolve the
abundances and $\mathcal E$ between hydrodynamic steps.  The resulting
changes are either driven by the hydrodynamic motion, and thus occur on the
same timescale, or are due to neutronization, which occurs on a slow
timescale compared to the hydrodynamics.  After a hydrodynamic step, the
values of $\mathcal E_n$, $\rho$, $T_n$ and $\{X'_i\}_n$ are available at each
grid point ($n$ labels the timestep).
The $q_n$ and $Y_{e,n}$ are calculated from $\{X'_i\}_n$, accounting for how
mixing changes these quantities.
This state is slightly out of NSE as a
consequence of the hydrodynamic evolution and a time interval $\Delta t$ has
passed during which some neutronization and neutrino loss has occurred.
We calculate $\dot Y_e$ and $\epsilon_\nu$ from tables using $\rho_n$, $T_n$,
and $Y_{e,n}$, and then set $Y_{e,n+1}= Y_{e,n}+\dot Y_e\Delta t$.  These
functions are not expected to vary steeply in these variables compared to the
amount they will change in a single hydrodynamical step.
We
then find $T_{n+1}$ by solving
\begin{eqnarray}
\label{eq:nseenergy}
\nonumber
\mathcal E[T_{n+1}, Y_{e,n+1}, \bar A(T_{n+1}, Y_{e,n+1})]
&-&q(T_{n+1}, Y_{e,n+1}) = \\ 
\mathcal E_n - q_n +
\Delta t[ -\dot Y_e N_A(&m_e&-m_n+m_p)c^2 -\epsilon_\nu ],
\end{eqnarray}
at fixed $\rho$.  Both $\bar A$ and $q$ are tabulated from the NSE
calculation using our full set of nuclear species.
Finally this temperature is used to calculate $\mathcal
E_{n+1}$, $q_{n+1}$ and $\bar A_{n+1}$, and the procedure described above is
used to convert $Y_{e,n+1}$, $\bar A_{n+1}$, and $q_{n+1}$ to
$\{X'_i\}_{n+1}$.

\section{One-dimensional Flame Model Results}
\label{sec:results}

In order to exercise the energy deposition and neutronization methods
described, we performed a series of one-dimensional model flame simulations
at a variety of densities.  In each of these a flame was propagated from left
to right through fuel consisting of equal parts by mass of \carbon\ and
\oxygen. In these simulations the input flame speeds were laminar
flame speeds (eq.~[\ref{eq:laminar}]); turbulent flame speeds 
(eq.~[\ref{eq:turb}]) were not used.
A reflecting boundary condition was used on the left, where the
flame is ignited,  and outflow on the right, so that fuel is free to
move off the grid as the volume of (less dense) ash increases.  A flame is
started by setting $\phi_1=1$ in a small region (with a smoothed edge)
adjacent to the left reflecting boundary and depositing the appropriate
energy locally. The 
Figs.~\ref{fig:1dr3e9}--\ref{fig:1dr5e7} show profiles of the abundances for
some of the surrogate nuclei and $Y_e$ at a time when the flame has
propagated most of the way across the domain.  Right panels show a detail of
the model flame region, which by definition is only several computational
zones wide.

\begin{figure}
\includegraphics[width=\hsize]{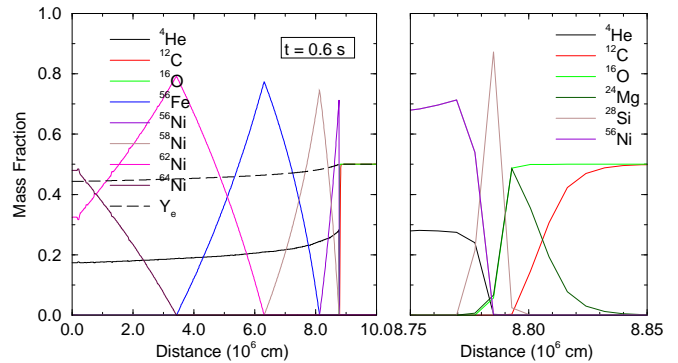}
\caption{
\label{fig:1dr3e9}
Abundances from a simulation of a model flame passing through 
a 50\%/50\% by mass mixture of C and O at an initial density of
$3 \times 10^{9} \gcc$. The left panel presents the 
abundances of selected surrogate nuclei on the entire simulation domain
at $t = 0.6 \mathrm{s}$. Also shown in the left panel is the electron fraction.
The right panel shows the detail of the thick flame
with another set of surrogate nuclei.
}
\end{figure}
\begin{figure}
\includegraphics[width=\hsize]{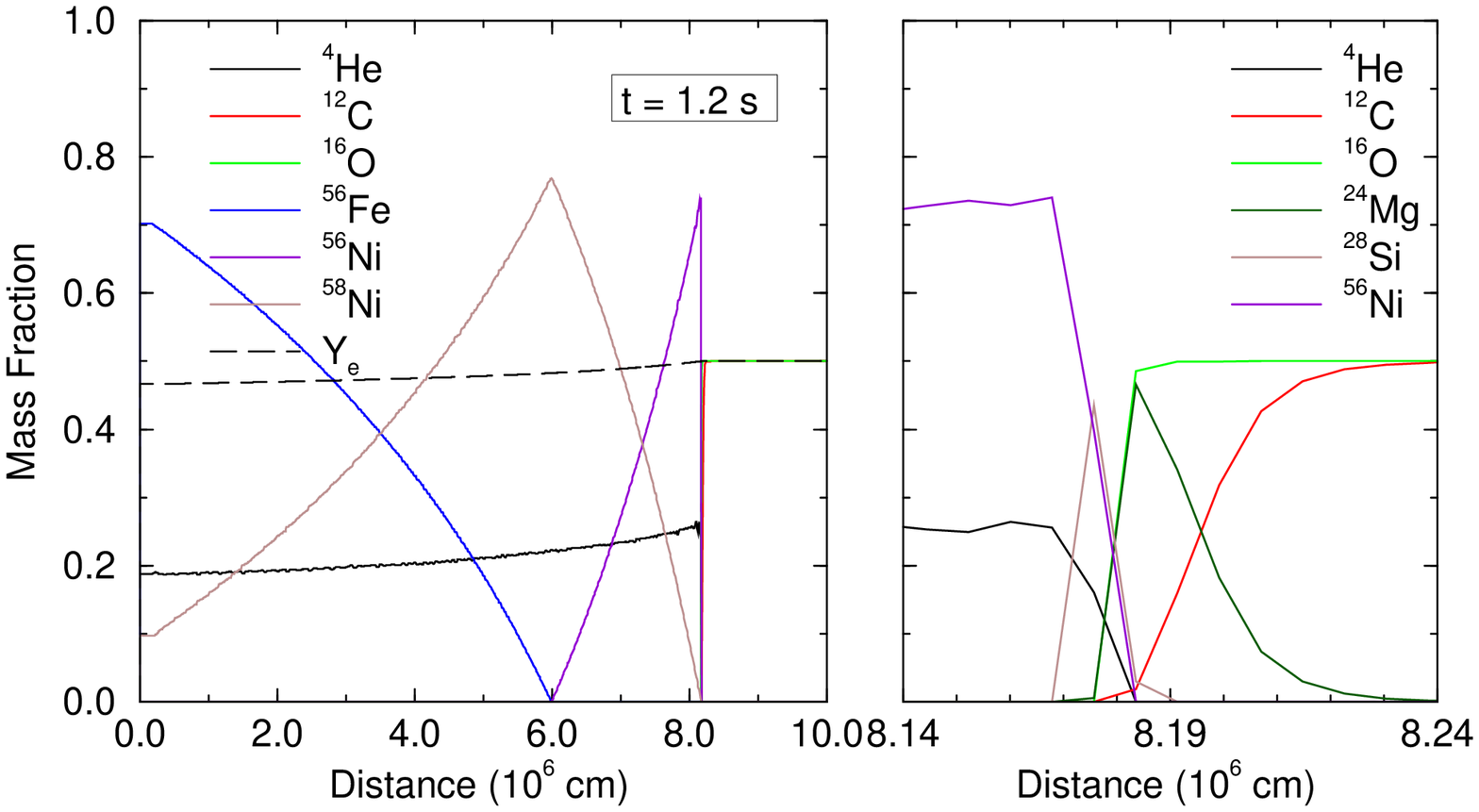}
\caption{
\label{fig:1dr1e9}
Abundances from a simulation of a model flame passing through 
a 50\%/50\% by mass mixture of C and O at an initial density of
$10^{9} \gcc$. The left panel presents the 
abundances of selected surrogate nuclei on the entire simulation domain
at $t = 1.2 \mathrm{s}$. Also shown in the left panel is the electron fraction.
The right panel shows the detail of the thick flame
with another set of surrogate nuclei. 
}
\end{figure}

Fig.s~\ref{fig:1dr3e9}--\ref{fig:1dr1e8} shows the results of a model
flames propagating across a $10^7 \mathrm{cm}$ domain with fuel at densities
of $3 \times 10^{9}$, $10^{9}$, and $10^{8} \gcc$, respectively. The elapsed
times of the simulations are $t = 0.6$, $1.2$ and $6.0\ \mathrm{s}$,
respectively, and the simulations were performed on a uniform mesh of 1280
points.  The material furthest to the left has been burned for the longest
time, and therefore has undergone the most neutronization.  Moving to the
right, ash is younger closer to the propagating flame.
The surrogate nuclear set works as expected, with \helium\ and consecutively
more neutron rich Fe-peak elements representing the NSE state.  The detailed
flame structure is also as expected, with the \carbon\ burning being followed
very closely (within a few zones) by conversion to \silicon\ representing NSQE
and that on to NSE.
The density dependence of the weak reactions is quite evident.
At $3\times 10^9\gcc$,  $Y_e=0.44$ is reached in only 0.6 s, though at
$10^9\gcc$, even after 1.2 s $Y_e$ has only fallen to about 0.47.
We performed an additional check on the neutronization by running a
self-heating calculation using the 200 nuclide network with weak reactions.
Comparing the degree of neutronization, $0.5-Y_e$, we found that the
self-heating calculation at $10^9\gcc$ differed from the leftmost zone of a
flame model simulation at initial density of $1.2\times 10^{9}\gcc$
(resulting in an ash density of $10^9\gcc$) by 2.6\% and 0.7\% at 0.6 and 1.2
s respectively.
Finally, the speed of the flame in the computational cell due to the creation
of the expanded ash is related to the laminar flame speed by $v =S_{\rm
laminar}\rho_{\rm fuel}/\rho_{\rm ash}$, and the test simulations reproduce this
behavior satisfactorily.


\begin{figure}
\includegraphics[width=\hsize]{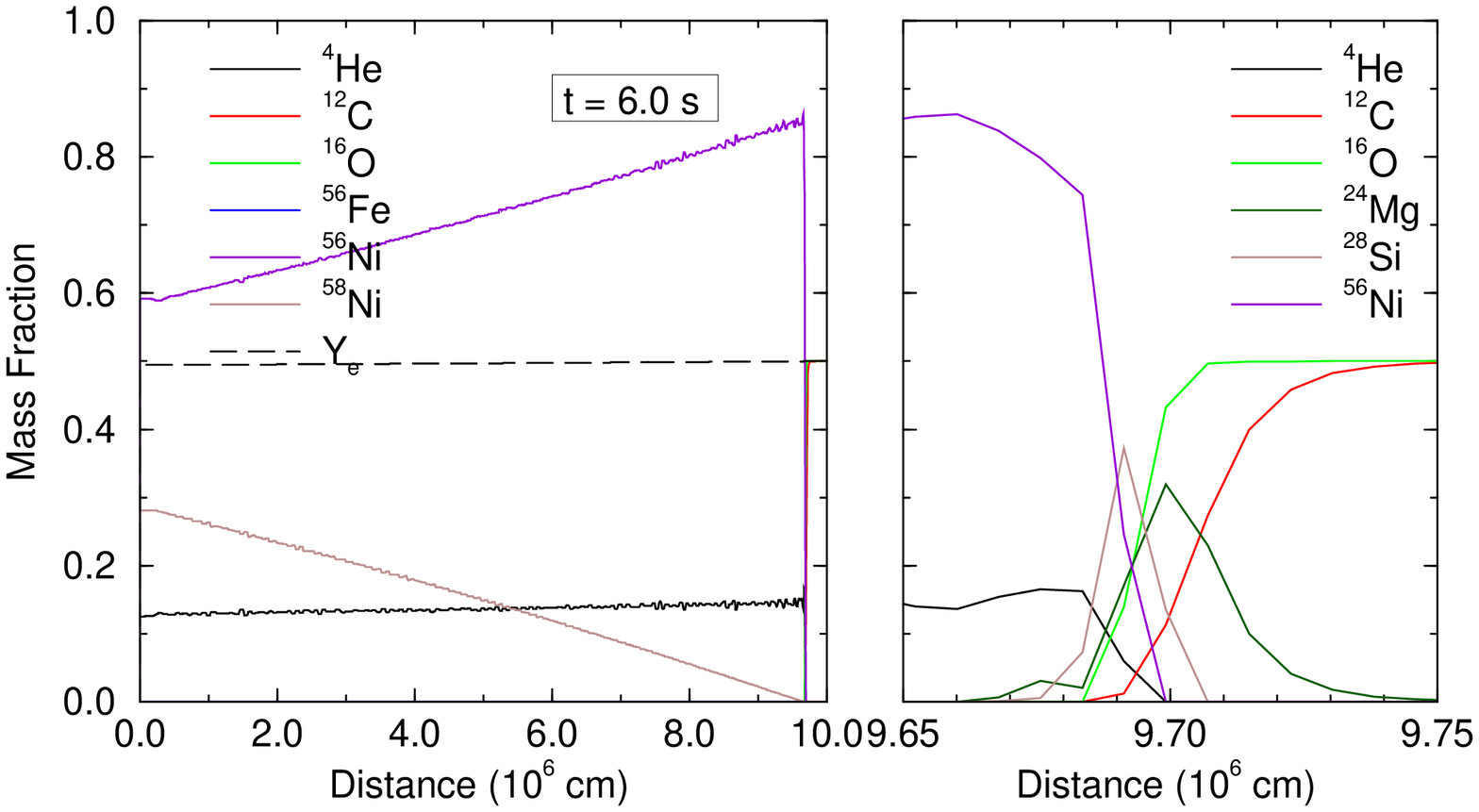}
\caption{
\label{fig:1dr1e8}
Abundances from a simulation of a model flame passing through 
an equal mass mixture of C and O at an initial density of
$1.0 \times 10^{8} \gcc$. The left panel presents the 
abundances of selected surrogate nuclei on the entire simulation domain
at $t = 6.0 \mathrm{s}$ Also shown in the left panel is the electron fraction.
The right panel shows the detail of the thick flame
with another set of surrogate nuclei. 
}
\end{figure}

Fig.~\ref{fig:1dr5e7} shows the result of a model flame propagating across
a larger  $8 \times 10^7 \mathrm{cm}$ domain with fuel at a density of 
$5.0 \times 10^{7} \gcc$. As with the high density simulations,
the simulation was performed on a uniform mesh of 1280 points.
The slower neutronization time scales with respect to burning
necessitated using the larger simulation domain to allow evolution of the
flame for a longer time, 16.0 s. In this case, the neutronization
following the flame is similar to the higher density flames although
the process of neutronization is much slower.
The detail of the flame in this case shows a broader \silicon\ region,
covering more than six computational zones,
indicating that at this low density the flame model is beginning to resolve
the NSE timescale.


\begin{figure}
\includegraphics[width=\hsize]{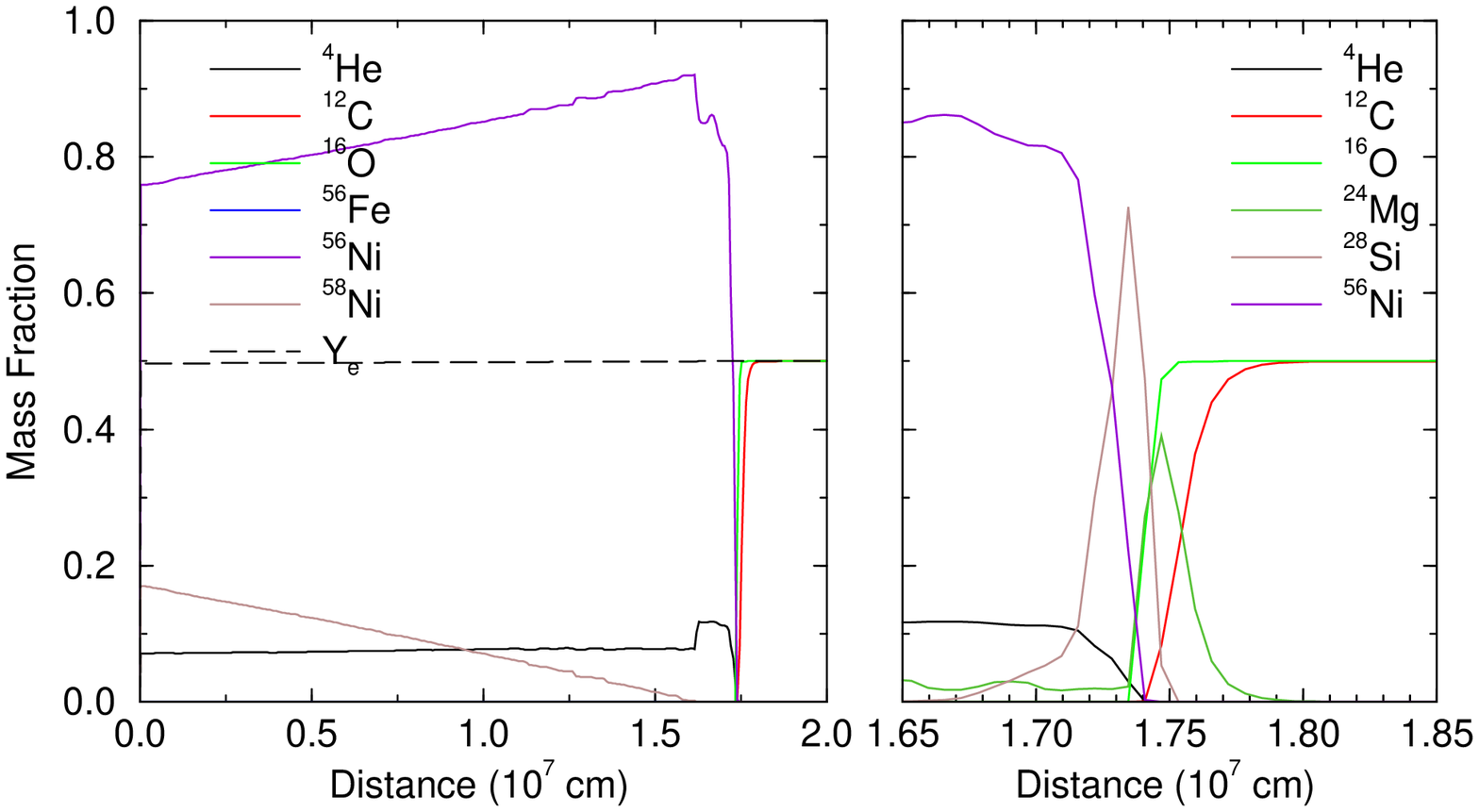}
\caption{
\label{fig:1dr5e7}
Abundances from a simulation of a model flame passing through 
an equal mass mixture of C and O at an initial density of
$5.0 \times 10^{7} \gcc$. The left panel presents the 
abundances of selected surrogate nuclei on the entire simulation domain
at $t = 16.0 \mathrm{s}$ Also shown in the left panel is the electron fraction.
The right panel shows the detail of the thick flame with another set of 
surrogate nuclei.
}
\end{figure}

In addition to these results, we performed a variety of test simulations
including model flames propagating away from an isolated ignition point in
the middle of the domain.  A similar calculation in which the fuel was moving
at a constant speed across the domain gave identical abundance profiles. The
results of these tests are shown in Figs.~\ref{fig:two_way} and 
\ref{fig:two_way_lag}. In both cases, the high degree of symmetry 
indicates that the direction in which the flame propagates from the 
ignition point does not affect the solution.  The simulation presented in 
Fig.~\ref{fig:two_way_lag} tested the ability of the code to correctly
propagate the flame in both directions as fuel moves across the grid
and indicates a satisfactory result.
\begin{figure}
\includegraphics[width=\hsize]{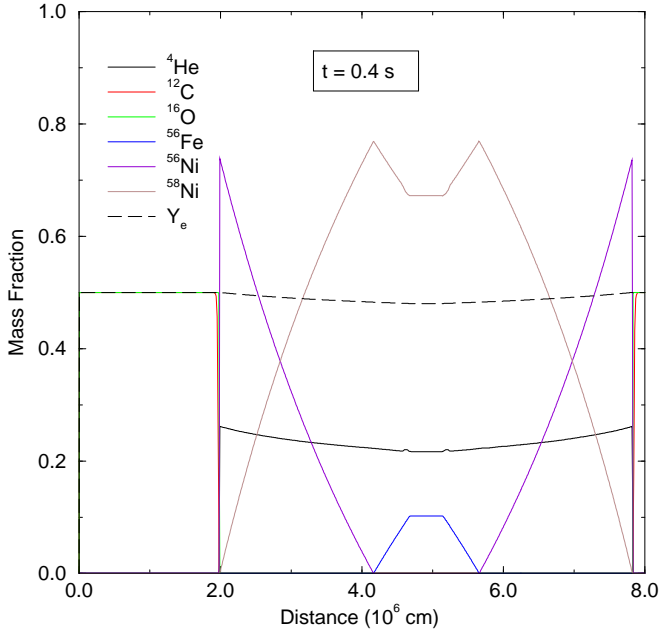}
\caption{
\label{fig:two_way}
Abundances and electron fraction from a simulation of a model flame passing 
through an equal mass mixture of C and O at an initial density of
$1.0 \times 10^{9} \gcc$. In this case, the ``match head" was placed
at the center of the domain so that the flame propagated to the left and 
right. The results may be compared to those of Fig.~\ref{fig:1dr1e9}, and
the obvious symmetry indicates the flame is correctly propagating in both
directions.
}
\end{figure}
\begin{figure}
\includegraphics[width=\hsize]{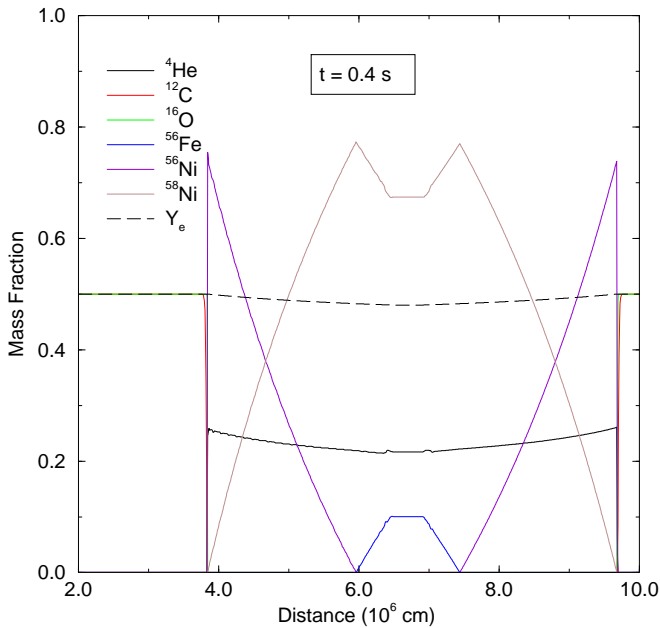}
\caption{
\label{fig:two_way_lag}
Abundances and electron fraction from a simulation of a model flame passing 
through an equal mass mixture of C and O at an initial density of
$1.0 \times 10^{9} \gcc$. In this case, the ``match head" was placed
at the center of the domain so that the flame propagated to the left and 
right while the material on the grid slowly flowed to the right. The results may be 
compared to those of Fig.~\ref{fig:two_way}, the principal difference being that
the symmetry is centered about $x = 6.6 \times 10^6$ cm. 
The obvious symmetry indicates the flame is correctly propagating in both
directions and that the slow propagation of the material across the 
simulation grid does not affect the results.
}
\end{figure}

\section{Conclusions}
\label{sec:conclusions}

We have extended the ADR flame model \citep{vladimirova+05} with multiple
burning stages and a reactive NSE post-flame state appropriate for
simulations of C/O deflagrations.  New calculations were performed to
calibrate flame structure, progress timescales, NSE properties and
neutronization.  This calibration involved detailed calculations of several
types, including direct flame simulation with a small nuclear network,
one-zone isochoric self-heating nuclear network calculations, and direct NSE
calculations.  We have described in detail how these calculation justify a
parameterized flame model, and what features and timescales must be
reproduced in a large-scale simulation.  By including Coulomb equation of
state corrections to detailed balance, we established self-consistency
between screened reaction network calculations and direct NSE calculations.
Our neutronization is calculated with up-to-date weak rates
\citep{langanke+01} convolved with our full NSE distribution.
We also developed a method of representing the precise binding energy and
$Y_e$ of the NSE state with a small set of nuclei that can then be easily
treated in a general multi-fluid code.

These improvements are necessary for several reasons. First, full-star models with embedded tracers for nucleosynthetic post-processing are now feasible. Following the passage of the flame, the NSE state evolves as large-scale fluid motions change its temperature and density. At higher densities, electron captures make the NSE state more neutron-rich and lower the pressure from degenerate electrons. Modeling both of these effects is critical for the tracers to capture a high-fidelity recording of the temperature-density history of the burn for subsequent nucleosynthesis calculations. Having a description of the NSE state that follows the post-explosion expansion and freeze-out is also important for evolving models far enough to compute lightcurves for contact with observations.

A second motivation for improving the flame model is to enable simulations of high-central density ignitions. Under these conditions, electron captures can play an important role in the energetics and yields of the explosion, and their influence must be accurately treated in the model flame. Observations are increasingly finding a diversity of Type Ia energetics and perhaps a diversity of progenitors \citep{Scannapieco2005The-Type-Ia-Sup,Mannucci2005Two-populations}. Connecting to observations demands that simulations of the explosion also operate under a diverse range of conditions. With this flame model, we can now simulate explosions from a wider range of progenitors. While this paper presents results for a \carbon/\oxygen\ mixture with a 1:1 mass ratio, the flame model is easily generalized to arbitrary \carbon:\oxygen\ ratios by varying $\Delta Q_{\mathrm{C}}$ and $X_{\mathrm{C}}^{0}$ (see eq.~[\ref{eq:flameenergy}] and following discussion).  The only limitation here is that $X_{\mathrm{C}}^{0}$ must be uniform throughout the white dwarf. Within this constraint of a spatially uniform initial composition, our generalized flame model is suitable for studies of thermonuclear deflagrations in white dwarfs under a range of ignition conditions.

A study of the hydrodynamic character of precisely how the three-stage burning
is integrated with the ADR flame and alternative ADR-type schemes is well
underway and, due to its relative complexity, will be published as a separate
work.  Here we have only presented 1-dimensional test calculations based on a
simple integration with the ADR flame model as it has been presented in the
literature \citep{khokhlov95}.  Future work will address the impact of many
3-dimensional flow characteristics such as flame front curvature, acoustic
behavior, flame front stability, and the effects of finite resolution.  Such a
hydrodynamical study is essential in order to understand WD deflagration
simulations performed with this flame-capturing technique, and to generally
understand the behavior of flames with a reactive ash such as this.

While improving the flame model is the motivation for this paper, our studies of explosive burning also clarify the importance of including Coulomb corrections to the equation of state and screening self-consistently.  Although these inclusions do not significantly affect the energy release 
and final temperature, they do decrease the timescales for reaching NSQE and NSE by 
factors of 2--3, with the most pronounced differences at high densities.
In the physical conditions under consideration, these Coulomb effects are
small but non-negligible, making it worth evaluating them accurately.
Since the effective turbulent flame speed is independent of the burning rate, it is less critical to compute the local flame speed
accurately.  We caution, however, that the determination of the effective flame speed is based on an assumption of \emph{steady} turbulent burning. It is therefore important to capture  the underlying nuclear physics with fidelity.

\acknowledgments
The authors thank Shimon Asida for his insight during discussions of
this work and acknowledge contributions from Tomasz Plewa.
The authors also thank Casey Dreier, an REU student at MSU, for
preliminary calculations that helped motivate this project.
The authors also thank Wolfgang Hillebrandt for previewing
and commenting on the manuscript.
This work is supported in part at the University of Chicago by the
U.S. Department of Energy under Grant B523820 to the ASC Alliances
Center for Astrophysical Flashes, and in part by
the National Science Foundation under Grant PHY 02-16783 for the
Frontier Center ``Joint Institute for Nuclear Astrophysics" (JINA). JWT acknowledges support 
from Argonne National Laboratory, which is operated under contract No. W-31-109-ENG-38 with 
the DOE.  EFB and ACC acknowledge support from the NSF grant AST-0507456.

\begin{appendix} 

\section{Plasma Coulomb Corrections for Nuclear Statistics}
\label{sec:coulomb_correction}

At the high densities that occur near the center in the initial phases of a
Type Ia Supernova, the screening of charged particle reactions by the
electron background can significantly enhance the reaction
rates.  This is, in fact, the same limit in which Coulomb\footnote{Note
that ``screening'' is used in some EOS literature to refer to only the dynamic
contribution of the electron background, not the effect of the
constant-density (static) electron background.  This leads to the
unfortunately worded but true
statement that in highly degenerate plasmas where the screening contribution
of the electrons to the EOS are negligible, the screening of reaction rates by
electrons can be quite strong.}
corrections to the
equation of state (EOS) become important for the statistical factors which
enter both a statistical equilibrium calculation and in determination of
reverse ratios for charged particle nuclear reactions.
In this appendix we outline our treatment of plasma Coulomb
corrections which preserves consistency between
screening-enhanced
reaction
network calculations used to obtain burning timescales and direct nuclear
statistical equilibrium (NSE) calculations used to follow the later evolution
of the NSE state.

The nuclear statistical equilibrium (NSE) state is defined by $\mu_i =
Z_i\mu_p+(A_i-Z_i)\mu_n$ where $\mu_i$ is the chemical potential (including
the nuclear rest mass energy)
of a given nuclide made up of $Z_i$ protons and $A_i-Z_i$
neutrons, $\mu_p$ and $\mu_n$ are the chemical potentials of the free
protons and neutrons.  We calculate the nuclear rest mass energy as $m_ic^2 =
Z_im_pc^2+(A_i-Z_i)m_nc^2-Q_i$, where $Q_i$ is the nuclear binding energy.
To obtain the correct NSE state in a dense plasma, $\mu_i$ must include
Coulomb
corrections.
The importance of these effects is measured by the Coulomb coupling
parameter $\Gamma=Z^2e^2/akT$ where $Z$ is the ion charge and $a=(4\pi
n/3)^{-1/3}$ is the average distance between ions, with $n$ representing the
ion number density.
It has been found \citep*{hansen+77,dewitt+96} that, to very good
approximation, the thermodynamic properties of multi-component plasmas can be
computed using the linear mixing rule, in which the correction to the
extensive Helmholtz free energy is summed for the constituents in proportion
to their number.  This results in a correction to $\mu_i=m_ic^2+\mu_i^{\rm
id}+\mu_i^{\rm C}$
for which we use the fitting form \citep{chabrier+98}
\begin{equation}
\label{eq:muiC}
\frac{\mu_i^{\rm C}}{kT} =  
A_1\left[\sqrt{\Gamma_i(A_2+\Gamma_i)} -
A_2\ln\left(\sqrt{\frac{\Gamma_i}{A_2}} +
\sqrt{1+\frac{\Gamma_i}{A_2}}\right)\right]
+ 2A_3\left[\sqrt{\Gamma_i}-\arctan\left(\sqrt{\Gamma_i}\right)\right]
,
\end{equation}
where $\Gamma_i = Z_i^{5/3}\Gamma_e$ is the ion-specific coulomb coupling
parameter with $\Gamma_e = e^2(4\pi n_e/3)^{1/3}/kT$, $A_1=-0.9052$,
$A_2=0.6322$, and $A_3=-\sqrt{3}/2-A_1/\sqrt{A_2}$.  This form is
accurate for both weakly ($\Gamma_i<0.1$) and strongly ($1\le\Gamma_i\lesssim160$)
coupled liquids, an important feature due to the wide variety of $Z_i$
present in an NSE calculation.  We have also included temperature dependent
nuclear partition functions \citep{rauscher+00} in $\mu_i^{\rm id}$.  More complicated
prescriptions for the multi-component plasma free energy have been explored
\citep{nadyozhin+05}, but the linear mixing rule appears sufficient for
our purposes of accomplishing correct energetics and hydrodynamics at
moderate $\Gamma$.
The abundances in NSE can now be found directly by using the equality
of chemical potentials (detailed balance) to write $X_i$ in terms of the
ideal part of the free nucleons' chemical potentials,
\begin{equation}
\label{saha}
X_i = \frac{m_i}{\rho}g_i \Big(\frac{2\pi m_i kT}{h^2}\Big)^{3/2}
\exp{\Big[\frac{Z_i \mu_p^{\rm id}+N_i \mu_n^{\rm id} + Q_i-\mu_i^C+Z_i \mu_p^C}{kT}\Big]}.
\end{equation}
To solve the system,
we substitute (\ref{saha}) into the constraint equations
\begin{eqnarray}
\label{massconstraint}
\sum_i X_i -1&=&0\\
\label{chargeconstraint}
\sum_i \Big[ (Y_e-1)Z_i+Y_e(A_i-Z_i) \Big]\frac{ X_i}{m_i}&=& 0
\end{eqnarray}
and solve numerically for $\mu_p^{\rm id}$ and $\mu_n^{\rm id}$ using a
Newton-Raphson method.

For charged particle reactions, rates are measured or
calculated in one direction and the reverse rate is then calculated
based on reciprocity and the species ratio in the NSE state (see e.g.\
\citealt{fowler+67}).  That is, for
the reaction $i(j,k)l$
\begin{equation}
\langle \sigma v\rangle_r = \left(\frac{n_in_j}{n_kn_l}\right)_{\rm NSE}
\langle\sigma v \rangle_f,
\end{equation}
where $\langle\sigma v\rangle$ denotes the product of the cross section and
velocity averaged over the appropriate distribution, for which we use
tabulated values from the REACLIB database~\citep{thielemann+86,rauscher+00}
and the $n_i$ denote the
number density of various particles.  This relation continues to hold when
plasma Coulomb corrections are important ($\Gamma_i\gtrsim0.1$) and thus the
rate enhancement factors for the forward and reverse rates must obey a
relationship which can be derived from the NSE condition.  We have found that
traditional screening factors (e.g.\ \citealt{wallace+82}) do not
inherently satisfy this relationship.  Even if they did, differences in the
fitting form
adopted for the plasma Coulomb corrections would still lead to inconsistency.
Therefore we have chosen to calculate the reverse ratios explicitly using the
same fitting form, (\ref{eq:muiC}), as for the NSE calculation.  This leads
to the relations
\begin{eqnarray}
\label{eq:inv}
\!\!\!\! \frac{\langle\sigma v\rangle_r}{\langle\sigma v\rangle_f} \! &=&
\! \frac{g_ig_j}{g_kg_l}
\left(\frac{m_im_j}{m_km_l}\right)^{3/2}
\!\!\! \exp\left(\frac{Q_i+Q_j-Q_k-Q_l}{kT}\right)
\exp\left(\frac{-\mu_i^{\rm C}-\mu_j^{\rm C}+\mu_k^{\rm C}+\mu_l^{\rm C}}{kT}\right)\\
\label{eq:invgam}
\!\!\!\! \frac{\lambda_\gamma}{\langle\sigma v\rangle_f}\! &=&
\! \frac{g_ig_j}{g_l}
\left(\frac{m_im_j}{m_k}\right)^{3/2}
\!\! \left(\frac{2\pi kT}{h^2}\right)^{3/2}
\!\!\! \exp\left(\frac{Q_i+Q_j-Q_l}{kT}\right)
\exp\left(\frac{-\mu_i^{\rm C}-\mu_j^{\rm C}+\mu_l^{\rm C}}{kT}\right)
\end{eqnarray}
for $i(j,k)l$ and $i(j,\gamma)l$ reactions respectively,
where $g_i$ are the temperature dependent nuclear partition functions, and
$m_i$ are the nuclear masses.  The forward rates themselves are found by
applying screening factors from \citet{wallace+82} to the rates
tabulated in the REACLIB database.

Although this formalism establishes consistency between a reaction network
calculation and a direct NSE calculation, some ambiguity remains.  A favored
reaction direction must be chosen and the reverse rate computed from its
screened rate.  The choice is apparent in the case of photodisintegrating
reverse reactions, but in the case of $(\alpha,p)$ reactions, for example,
the choice is less clear.  In practice one direction typically has a superior
calculation or measurement of the nuclear cross section, or one direction is
not actually known at all.  As shown below, it is not necessary to reconcile
this ambiguity in the current work, but it is enlightening to examine the
relationship between the screening corrections to the rates and the reverse
ratio presented in (\ref{eq:inv}) and (\ref{eq:invgam}).  The screening
enhancement $f=\exp(H)$ is given by \citet{wallace+82} (see also
\citealt{ogata+93} for a similar expression),
\begin{equation}
H= C -\frac{\tau}{3}\left(\frac{5}{32}b^3-0.014b^4 -0.128b^5\right)
-\Gamma(0.0055b^4-0.0098b^5+0.0048b^6)
\end{equation}
where $b=3\Gamma_{\rm eff}/\tau_{12}$,
$\tau_{12}=[(27\pi^2/4)2\hat AZ_1^2Z_2^2
e^4/N_AkTh^2]^{1/3}=4.2(Z_1^2Z_2^2\hat A_{12}/T_9)^{1/3}$, and $\hat
A=A_1A_2/(A_1+A_2)$ are parameters related to the quantum contributions 
and the classical plasma contributions are contained in
\begin{equation}
C=f_{\rm ex}(Z_1^{5/3}\Gamma_e)+f_{\rm ex}(Z_2^{5/3}\Gamma_e)-f_{\rm
ex}[(Z_1+Z_2)^{5/3}\Gamma_e]
\end{equation}
where $f_{\rm ex}(\Gamma_i)=\mu_i^{\rm C}/kT$ is the excess free energy due
to Coulomb corrections.  For the situations under consideration here,
$\Gamma\lesssim10$ and $T_9\sim 4$-9, so that $b$ is fairly small.  In this
situation the classical correction, $C$, dominates, so that for an
$i(j,k)l$ reaction $\langle\sigma v\rangle_f=
\langle\sigma v\rangle^{\rm B}_f\exp(C_{ij})$,
and where B denotes the bare (unscreened) reaction rate.  We thus find
that because $Z_i+Z_j=Z_k+Z_l$, the reverse ratio equation (\ref{eq:inv})
changes the $C_{ij}$ multiplying the forward rate into $C_{kl}$, giving the
the proper screening enhancement for the reverse rate.  This does not address
differences in the fitting form adopted for $f_{\rm ex}(\Gamma_i)$, which
should be small by definition, but which we have avoided by using the same
$f_{\rm ex}(\Gamma_i)$ in the reverse ratios and the NSE calculation.

\end{appendix}  

\clearpage

\begin{landscape}  

\begin{deluxetable}{c|cccc|cccc|cccc}[p]
\tabletypesize{\scriptsize}
\tablecaption{Temperature, energy release, H and He mass fractions for
isochoric, self-heating network calculations with three different networks. 
}
\tablehead{
\colhead{}  &  \multicolumn{4}{c}{P200} & \multicolumn{4}{c}{aprox19}  &  \multicolumn{4}{c}{torch47} \\
\colhead{log($\rho$)} &   \colhead{$T_{\rm NSE}$} & \colhead{$\Delta$ Q} & \colhead{$X_p$} & \colhead{$X_\alpha$} 
& \colhead{$T_{\rm NSE}$} & \colhead{$\Delta$ Q} & \colhead{$X_p$} & \colhead{$X_\alpha$} &  
\colhead{$T_{\rm NSE}$} & \colhead{$\Delta$ Q} & \colhead{$X_p$} & \colhead{$X_\alpha$} \\
\colhead{($\gcc$)} & \colhead{({$10^9$} K)} & \colhead{($10^{17} \egg $)} & 
\colhead{ (\%) } & \colhead{ (\%) } & \colhead{({$10^9$} K)} & \colhead{($10^{17} \egg$)} & 
\colhead{ (\%) } & \colhead{ (\%) } & \colhead{({$10^9$} K)} & \colhead{($10^{17} \egg$)} & \colhead{ (\%) } & \colhead{ (\%) }
}
\startdata
 7.0 & 4.56 &  7.18 & 0.847 &  0.802 &  4.54  & 6.78  &  0.516 &  0.334 &  4.46  &  7.53  &  0.214 &  0.867\\
 7.1 & 4.75 &  6.95 &  1.09 &  1.30  &  4.77  & 6.67  &  0.776 &  0.890 &  4.78  &  7.22  &  0.543 &  0.872\\
 7.2 & 4.95 &  6.69 &  1.34 &  1.97  &  4.96  & 6.42  &  1.02  &  1.55  &  4.99  &  7.04  &  0.693 &  1.86\\
 7.3 & 5.13 &  6.40 &  1.57 &  2.81  &  5.15  & 6.20  &  1.27  &  2.50  &  5.20  &  6.76  &  0.834 &  3.11\\
 7.4 & 5.32 &  6.14 &  1.77 &  3.52  &  5.33  & 5.98  &  1.52  &  3.99  &  5.41  &  6.47  &  0.947 &  4.47\\
 7.5 & 5.50 &  5.85 &  1.96 &  4.66  &  5.52  & 5.71  &  1.70  &  5.28  &  5.63  &  6.13  &  1.04  &  6.17\\
 7.6 & 5.68 &  5.57 &  2.12 &  5.86  &  5.68  & 5.42  &  1.85  &  6.83  &  5.82  &  5.87  &  1.10  &  7.52\\
 7.7 & 5.85 &  5.30 &  2.25 &  7.14  &  5.83  & 5.20  &  1.95  &  8.30  &  5.99  &  5.65  &  1.14  &  8.72\\
 7.8 & 6.01 &  5.05 &  2.35 &  8.34  &  5.88  & 4.92  &  2.03  &  9.72  &  6.17  &  5.44  &  1.16  &  9.92\\
 7.9 & 6.17 &  4.82 &  2.44 &  9.45  &  5.94  & 4.76  &  2.35  &  11.0  &  6.35  &  5.23  &  1.18  &  11.1\\
 8.0 & 6.33 &  4.62 &  2.51 &  10.5  &  6.40  & 4.56  &  2.37  &  12.6  &  6.53  &  5.02  &  1.20  &  12.3\\
 8.1 & 6.50 &  4.43 &  2.57 &  11.4  &  6.47  & 4.40  &  2.45  &  13.6  &  6.74  &  4.81  &  1.21  &  13.6\\
 8.2 & 6.66 &  4.27 &  2.62 &  12.3  &  6.61  & 4.25  &  2.46  &  14.5  &  6.90  &  4.67  &  1.22  &  14.5\\
 8.3 & 6.82 &  4.12 &  2.66 &  13.0  &  6.76  & 4.11  &  2.46  &  15.5  &  7.07  &  4.51  &  1.23  &  15.4\\
 8.4 & 6.99 &  3.98 &  2.69 &  13.7  &  6.91  & 4.00  &  2.46  &  16.3  &  7.25  &  4.38  &  1.23  &  16.2\\
 8.5 & 7.15 &  3.86 &  2.73 &  14.3  &  7.07  & 3.88  &  2.46  &  17.3  &  7.44  &  4.22  &  1.24  &  17.0\\
 8.6 & 7.33 &  3.76 &  2.76 &  14.8  &  7.22  & 3.80  &  2.46  &  18.1  &  7.63  &  4.10  &  1.24  &  17.6\\
 8.7 & 7.50 &  3.66 &  2.79 &  15.3  &  7.38  & 3.71  &  2.50  &  18.8  &  7.82  &  4.04  &  1.24  &  18.2\\
 8.8 & 7.68 &  3.57 &  2.81 &  15.7  &  7.55  & 3.64  &  2.54  &  20.9  &  8.03  &  3.92  &  1.25  &  18.6\\
 8.9 & 7.87 &  3.49 &  2.84 &  16.1  &  7.73  & 3.58  &  2.53  &  22.5  &  8.23  &  3.85  &  1.25  &  19.0\\
 9.0 & 8.06 &  3.42 &  2.86 &  16.4  &  8.55  & 3.50  &  2.58  &  23.8  &  8.45  &  3.80  &  1.27  &  19.3\\
 9.1 & 8.26 &  3.35 &  2.89 &  16.7  &  8.61  & 3.47  &  2.60  &  25.1  &  8.71  &  3.73  &  1.28  &  19.5\\
 9.2 & 8.47 &  3.29 &  2.91 &  16.9  &  8.81  & 3.42  &  2.61  &  25.3  &  8.92  &  3.68  &  1.28  &  19.7\\
 9.3 & 8.69 &  3.24 &  2.94 &  17.1  &  8.89  & 3.39  &  2.70  &  25.3  &  9.15  &  3.65  &  1.29  &  19.7\\
 9.4 & 8.92 &  3.19 &  2.96 &  17.3  &  9.35  & 3.36  &  2.70  &  25.4  &  9.39  &  3.60  &  1.31  &  19.7\\
 9.5 & 9.16 &  3.15 &  2.98 &  17.4  &  9.56  & 3.32  &  2.70  &  25.6  &  9.67  &  3.55  &  1.32  &  19.6\\
 9.6 & 9.40 &  3.11 &  3.01 &  17.5  &  9.83  & 3.28  &  2.78  &  25.7  &  9.92  &  3.50  &  1.35  &  19.7\\
 9.7 & 9.66 &  3.07 &  3.03 &  17.5  &  10.0  & 3.26  &  2.78  &  25.7  &  10.2  &  3.47  &  1.37  &  19.8\\
 9.8 & 9.94 &  3.04 &  3.05 &  17.6  &  10.4  & 3.26  &  2.80  &  25.7  &  10.4  &  3.45  &  1.41  &  20.1\\
 9.9 & 10.2 &  3.01 &  3.07 &  17.6  &  10.6  & 3.24  &  2.80  &  25.9  &  10.7  &  3.42  &  1.45  &  21.1\\
10.0 & 10.5 &  2.98 &  3.09 &  17.6  &  11.1  & 3.22  &  2.80  &  25.9  &  11.1  &  3.39  &  1.67  &  21.1 \\
\enddata
\tablecomments{P200: the 200 nuclide network described in this paper; 
\textit{aprox19}: the approximated 19 nuclide network~\citep{wzw1978,timmes+99}; 
\textit{torch47}: the 47 nuclide network~\citep{timmes+99}. }
\label{tab:big}
\end{deluxetable}
\clearpage

\end{landscape} 

\begin{deluxetable*}{ccccccc}
\tabletypesize{\scriptsize}
\tablewidth{0pt}
\tablecaption{Temperature, energy release, H and He mass fractions for
isobaric, self-heating network calculations with 200 nuclide network. 
}
\tablehead{ 
\colhead{log(P)} & \colhead{log($\rho_{\rm fuel}$)} & \colhead{log($\rho_{\rm ash}$)} & \colhead{$T_{\rm NSE}$} & \colhead{$\Delta$ Q} & \colhead{$X_p$} & \colhead{$X_\alpha$} \\ 
\colhead{(dyne/cm$^2$)} & \colhead{(g/cm$^3$)} & \colhead{(g/cm$^3$)} & \colhead{({$10^9$} K)} & \colhead{($10^{17} \egg $)}  & \colhead{(\%)}  & \colhead{(\%)}   
} 
\startdata
24.0  &  6.93 & 6.54 &   3.43 &   7.82 &  0.0564 &  0.00785 \\
24.1  &  7.02 & 6.63 &   3.60 &   7.80 &  0.0932 &  0.0181  \\
24.2  &  7.10 & 6.72 &   3.77 &   7.75 &  0.149  &  0.0397  \\
24.3  &  7.18 & 6.82 &   3.94 &   7.69 &  0.230  &  0.0826  \\
24.4  &  7.26 & 6.91 &   4.13 &   7.61 &  0.343  &  0.163   \\
24.5  &  7.34 & 7.00 &   4.31 &   7.49 &  0.491  &  0.303   \\
24.6  &  7.41 & 7.09 &   4.50 &   7.33 &  0.674  &  0.533  \\
24.7  &  7.49 & 7.18 &   4.70 &   7.14 &  0.886  &  0.882  \\
24.8  &  7.57 & 7.28 &   4.89 &   6.91 &  1.12   &  0.138  \\
24.9  &  7.64 & 7.37 &   5.07 &   6.66 &  1.35   &  2.03   \\
25.0  &  7.72 & 7.46 &   5.26 &   6.40 &  1.57   &  2.83   \\
25.1  &  7.80 & 7.55 &   5.44 &   6.12 &  1.77   &  3.77   \\
25.2  &  7.87 & 7.64 &   5.61 &   5.85 &  1.94   &  4.79   \\
25.3  &  7.95 & 7.74 &   5.78 &   5.60 &  2.09   &  5.87   \\
25.4  &  8.02 & 7.83 &   5.94 &   5.35 &  2.21   &  6.95   \\
25.5  &  8.10 & 7.91 &   6.10 &   5.12 &  2.31   &  8.01   \\
25.6  &  8.18 & 8.00 &   6.25 &   4.91 &  2.40   &  9.01   \\
25.7  &  8.25 & 8.09 &   6.40 &   4.72 &  2.47   &  9.95   \\
25.8  &  8.33 & 8.17 &   6.55 &   4.54 &  2.53   &  10.8   \\
25.9  &  8.40 & 8.26 &   6.70 &   4.38 &  2.58   &  11.6   \\
26.0  &  8.48 & 8.34 &   6.84 &   4.24 &  2.62   &  12.4   \\
26.1  &  8.55 & 8.43 &   6.99 &   4.11 &  2.66   &  13.0   \\
26.2  &  8.63 & 8.51 &   7.13 &   3.99 &  2.69   &  13.6   \\
26.3  &  8.70 & 8.59 &   7.28 &   3.88 &  2.72   &  14.2   \\
26.4  &  8.78 & 8.68 &   7.43 &   3.78 &  2.75   &  14.6   \\
26.5  &  8.85 & 8.76 &   7.58 &   3.69 &  2.77   &  15.1   \\
26.6  &  8.93 & 8.84 &   7.73 &   3.61 &  2.80   &  15.5   \\
26.7  &  9.00 & 8.92 &   7.89 &   3.54 &  2.82   &  15.8   \\
26.8  &  9.08 & 9.00 &   8.04 &   3.47 &  2.84   &  16.1   \\
26.9  &  9.15 & 9.08 &   8.21 &   3.41 &  2.86   &  16.4   \\
27.0  &  9.23 & 9.16 &   8.37 &   3.36 &  2.88   &  16.6   \\
27.1  &  9.31 & 9.24 &   8.54 &   3.31 &  2.90   &  16.8   \\
27.2  &  9.38 & 9.32 &   8.72 &   3.26 &  2.92   &  17.0   \\
27.3  &  9.46 & 9.40 &   8.90 &   3.23 &  2.93   &  17.1   \\
27.4  &  9.53 & 9.47 &   9.08 &   3.19 &  2.95   &  17.2   \\
27.5  &  9.61 & 9.55 &   9.27 &   3.16 &  2.97   &  17.2   \\
27.6  &  9.68 & 9.63 &   9.47 &   3.14 &  2.98   &  17.3   \\
27.7  &  9.76 & 9.71 &   9.67 &   3.11 &  3.00   &  17.3   \\
27.8  &  9.83 & 9.79 &   9.88 &   3.09 &  3.02   &  17.3   \\
27.9  &  9.91 & 9.86 &   10.1 &   3.08 &  3.03   &  17.2   \\
28.0  &  9.98 & 9.94 &   10.3 &   3.07 &  3.04   &  17.2   \\
28.1  &  10.1 & 10.0 &   10.5 &   3.06 &  3.06   &  17.2   \\
28.2  &  10.1 & 10.1 &   10.8 &   3.05 &  3.06   &  17.1   \\
28.3  &  10.2 & 10.2 &   11.0 &   3.05 &  3.06   &  17.0   \\
28.4  &  10.3 & 10.3 &   11.3 &   3.05 &  3.07   &  16.9   \\
28.5  &  10.4 & 10.3 &   11.5 &   3.05 &  3.08   &  16.7   \\
\enddata								 
\label{tab:nse_isobaric}						
\end{deluxetable*}							  
									  
\end{document}